\documentclass[12pt]{iopart}

\usepackage{graphicx}
\usepackage{bm}

\usepackage{placeins}

\usepackage[colorlinks=true]{hyperref}

\begin{document}

\def\t{\tilde}

\title{Radioscience simulations in General Relativity and in alternative theories of gravity}

\author{A.~Hees$^{1,2,3}$, B.~Lamine$^4$, S.~Reynaud$^4$, M.-T.~Jaekel$^5$, C.~Le~Poncin-Lafitte$^2$, V.~Lainey$^6$, A.~F\"uzfa$^3$, J.-M~Courty$^4$, V.~Dehant$^1$ and P.~Wolf$^2$}

\address{$^1$ Royal Observatory of Belgium,  Avenue Circulaire 3, 1180 Bruxelles, Belgium}
\address{$^2$ LNE-SYRTE, Observatoire de Paris, CNRS, UPMC,\\ Avenue de l'Observatoire 61, 75014 Paris}
\address{$^3$ Namur Center for Complex Systems (naXys), \\University of Namur (FUNDP), Belgium}
\address{$^4$ Laboratoire Kastler Brossel, \'Ecole Normale Sup\'erieure, CNRS, UPMC,\\ 75252 Paris Cedex 05, France }
\address{$^5$ Laboratoire de Physique Th\'eorique de l'\'Ecole Normale Sup\'erieure, CNRS, UPMC,\\ Rue Lhomond 24, 75005 Paris, France}
\address{$^6$ IMCCE, Observatoire de Paris, UMR 8028 du CNRS, UPMC, Universit\'e de Lille 1}
                                                                                            
\ead{aurelien.hees@oma.be}

\date{June, 2012}

\submitto{\CQG}

\begin{abstract}
  This paper deals with tests of General Relativity in the Solar System
  using tracking observables from planetary spacecraft. We present a
  new software that simulates the Range and Doppler signals resulting from a given
 space-time metric. This flexible approach allows one to perform
  simulations in General Relativity as well as in alternative metric
  theories of gravity. The outputs of this software provide templates of
  anomalous residuals that should show up in real data if the
  underlying theory of gravity is not General Relativity. Those
  templates can be used to give a rough
  estimation  of
  constraints on additional parameters entering alternative theory of
  gravity and also signatures that can be searched for in data from past or future
  space missions aiming at testing gravitational laws in the Solar
  System. As an application of the potentiality of this software, we
  present some simulations performed for Cassini-like mission in
  Post-Einsteinian Gravity and in the context of MOND External Field
  Effect. We derive signatures arising from these alternative
  theories of gravity and estimate expected amplitudes of the anomalous
  residuals.
\end{abstract}

\maketitle
\section{Introduction}
Testing General Relativity (GR) is a long standing and worthy effort
in the scientific community. From a theoretical point of view, the
different attempts to quantize gravity or to unify it with the other
fundamental interactions always predict deviations from GR. From an
observational point of view, cosmological data cannot be explained by
the combination of GR and the standard model of particles. In the most
accepted cosmological model, the so-called $\Lambda$CDM model, these
observations are explained by the presence of two puzzling
ingredients: Dark Matter and Dark Energy, representing respectively
about 22~\% and 74~\% of the Universe.  Until today, these two dark
components have not been observed directly. Therefore these
cosmological observations can be a hint that General Relativity may not be the correct theory of gravitation at
large scales.

Up to now, GR has passed all the critical tests in the different
situations where it has been tested. In the Solar System, the tests of
gravitation mainly rely on the parameterized Post-Newtonian (PPN)
formalism \cite{will:1993fk,will:2006cq} or on a search for a
deviation from the Newtonian potential of the Yukawa type (the
so-called fifth force search). In the PPN framework the space-time
metric is parameterized by 10 constant coefficients (the most important
ones being $\gamma$ that characterizes the spatial curvature and
$\beta$ that characterizes the non-linearity of the theory). The
current constraints on these coefficients are very stringent as
described in Will~\cite{will:2006cq}. For example, the $\gamma$
parameter is constrained at the level of $10^{-5}$ by the measurement
of the Shapiro delay using the Cassini
spacecraft~\cite{bertotti:2003uq} while the $\beta$ parameter is now
constrained at the level of $10^{-5}$ with planetary
ephemeris~\cite{fienga:2011qf}, Lunar Laser
Ranging~\cite{williams:2009ys} and with the tracking of Mars
orbiters~\cite{konopliv:2011dq}.  In the fifth force framework
(described
in~\cite{adelberger:2003uq,adelberger:2009fk,talmadge:1988uq,
  fischbach:1999ly}), the gravitation theory is very well tested at
almost all scales as can be seen from figure 31
of~\cite{konopliv:2011dq}. Nevertheless, there are still two main open
windows where potential deviations can be expected: at small scales in
the laboratory and at outer solar-system scales.  In this context, let
us mention the existence of an anomalous acceleration recorded on
Pioneer 10/11 probes during their flight to the outer solar
system~\cite{anderson:1998fk,anderson:2002uq,levy:2009vn} (for a
review, see~\cite{turyshev:2010kx} and references therein). If recent
publications~\cite{turyshev:2011fk,turyshev:2012fk} seem to indicate that part of the
secular anomaly could be accounted by thermal effects, other features
such as modulations call for another explanation.

Given the existing stringent constraints, there are strong motivations
to move forward in the search for deviations from GR in the Solar
System. First of all, to push for higher precision experiments remains
a valuable challenge. This is motivated by some scenarios of
alternative theories of gravity producing deviations smaller than the
current constraints. Let us mention in this context certain types
of tensor-scalar theories where the cosmological evolution exhibits an
attractor mechanism that attracts the theory towards
GR~\cite{damour:1993kx,damour:1993uq} or  chameleon
theories~\cite{khoury:2004fk,khoury:2004uq,hees:2012kx} where the deviations from
GR are hidden in the region of the Universe where the matter density
is high (as inside the solar system). Both of these alternative
theories of gravity produce a deviation of the $\gamma$ PPN parameter
smaller than the current constraint. There exist also open windows in
existing frameworks where deviations can be searched at very small
distances and at very large distances (in the outer Solar
System). Finally, it is useful to analyze experimental results in new
extended frameworks. Indeed, even if observations lie very close to GR
when analyzed within the PPN or fifth force framework, this does not
mean that this has to be true in any other framework. The existing
frameworks indeed cover a limited set of alternative theories of
gravity. A formalism based on a non-local Einstein field equation more
adapted for quantization has thus been developed in a series of recent
papers~\cite{jaekel:2005zr,jaekel:2005vn,jaekel:2006kx,jaekel:2006uq,reynaud:2007fk}.
In this extension of GR, the modification of the space-time metric in
the Solar System is phenomenologically described by two functions
depending on position. As is also the case in $f(R)$ theories exhibiting a chameleon
mechanism~\cite{faulkner:2007kx,capozziello:2008vn,de-felice:2010uq},
the PPN parameters are then promoted to functions depending on
position. A last example is given by the Standard Model Extension
(SME) framework where Lorentz symmetry is broken. In the gravitational
sector, SME is characterized by a metric which goes beyond the
existing framework~\cite{bailey:2006uq,bailey:2009fk}.

In this paper, we focus on Solar System tests of gravity. In this
context, the gravitational observations rely either on astrometric
observations (right ascension and declination), with the advantage of
very long measurement time, or on spacecraft tracking (radioscience
measurement : Range, Doppler, VLBI), with the advantage of very high
precision but limited time span. Here, we only consider radioscience
measurements based on Range and Doppler (future work will consider
angular measurements).

Our present work is part of a long term project aiming at allowing,
\emph{systematic} and \emph{versatile} scanning of data from solar
system observations (e.g. spacecraft Range, Doppler, VLBI, angular
measurements, astrometry, \dots) looking for possible violations of
the known laws of gravitation (General Relativity). By this we mean
that the same basic procedure can be systematically applied to all
different types of data, or their combination; and by versatile we
mean that the procedure is easily adapted to any alternative theory
that is tested for, provided the space-time metric of that theory is
available.

The basic procedure, when completed, will consist of the following steps:
\begin{enumerate}
	\item   Simulate the observables of a given physical situation (eg. an arc of spacecraft tracking from Earth) in the alternative theory. In this step, a simplified version of the physical situation can be considered: only the elements that can give rise to leading order deviations due to the alternative theory need to be simulated.
	\item  Analyze the resulting observables using the usual procedure in GR: fit a GR model on the data by adjusting the initial conditions of the different bodies and the parameters entering the modeling~\cite{zarrouati:1987fk}. The residuals of that analysis provide the incompressible signal that should be present in the residuals of the real data of the considered physical situation if the gravitation theory in the solar system is not GR but the one used in the first step. These residuals are called incompressible in the sense that this part of the signal can not be absorbed anymore by a fit of real or simulated (in an alternative theory of gravity) data using a GR model.
	\item  Analyze the real data using the usual procedure in GR (including all known systematic effects eg solar radiation, thermal effects, all gravitational perturbations, etc\dots). This analysis provides residuals of a fit of real data using a GR model.
	\item  Systematically search the residuals of step (iii) (obtained from real data) using the "template" obtained in step (ii). This can be done by optimal filtering, matched filtering or any other statistical method best adapted to the data and template. If the template is found with a S/N $>$ 1 the alternative theory considered is better supported by the data than GR. Depending on the S/N one can then consider the result significant or not, and start searching for systematic effects that might explain it (helped by the detailed signature of the template) and/or try other physical situations (other spacecraft, other types of observation, etc\dots) to confirm the result.
\end{enumerate}
In this paper we present a first version of the step (i) and a simplified version of the step (ii). The simulation software is built in the spirit that everything is computed from the space-time metric (computation of trajectories, clock behavior, light propagation). This software allows simulation of spacecraft Range and Doppler observations in any alternative theory for which a metric is available.  Note that in the first step only the leading order effects need to be considered, as we expect the modifications of second order effects (eg. solar radiation, thermal effects, gravitational perturbations, ...) by the alternative theory to be negligible with respect to the leading order. However, step (ii) needs to include all effects that could absorb some of the anomalous residuals by fitted parameters of the perturbations. Some care is required concerning the coherence between steps (i) and (ii), to ensure that no "false" signals are generated by perturbations included in step (ii) but not in (i). As an example, including perturbations by an additional planet in only step (ii) (but not (i)) will significantly modify the residuals (obtained from simulated data), but including it in both or neither gives rise to essentially the same residuals (cf. \sref{sec:simuPEG} for an explicit example using Jupiter). In the simplified version presented here the second step uses only the main features of a full analysis in GR (neglecting eg. solar radiation pressure, thermal effects, planetary gravitational fields, etc\dots).

Nonetheless this allows us to derive the general form of the expected templates for a given physical situation in alternative theory, although some of the features of those templates will be modified (absorbed by additional parameters) when carrying out a complete fit in the second step. Furthermore, we can compare the maximum amplitude of those templates to the typical rms noise of residuals from real spacecraft tracking, thereby giving a rough order of magnitude of the expected constraints when a full analysis of the type above is applied to existing data. On one hand the constraints in a full analysis will be more stringent than these rough orders of magnitude because optimum filtering will perform better than a simple comparison of the residual rms to the template amplitude. On the other hand they will be less stringent because some of the features of the signature will be absorbed by additional parameters that are not present in our simplified version of step (ii). Finally, for future missions our procedure will allow optimizing the mission characteristics (trajectories, periods of observation, maneuvers, etc\dots) for optimum tests of alternative theories by maximizing the signatures in the templates.   

The procedure presented above is significantly easier to apply (once the software for steps (i) and (ii) exists) and more versatile, than to write a complete data analysis software in an alternative theory. The main reason for that is that in steps (i) and (ii) a simplified situation is sufficient to obtain the correct template. It is not necessary to include all perturbing effects (solar radiation pressure, all planets and asteroids, thermal radiation, etc\dots) at this stage. The full analysis including all effects is carried out in step (iii) using existing software in GR. Furthermore residuals from step (iii) are in general more readily available and easier to handle than raw data from the observations.

\section{Outline} \label{sec:outline}
The method used to simulate Range and
Doppler observations directly from the space-time metric is
presented in detail in \sref{sec:num}. The definitions of the observables (Range and
Doppler) are given and the methodology used to simulate such
observables is presented step by step: the derivation and the
integration of the equations of motion, the behavior of the clocks
and the propagation of light in curved space-time. With this
software, it is possible to simulate radioscience signals in
alternative theories of gravity.

In \sref{sec:comp}, we explain the method used to analyze signals
obtained. In particular, the least-square fit of
the initial conditions is briefly described and an estimation of the
numerical accuracy of the whole process (simulation and fit procedure)
is presented.

\Sref{sec:results} presents the original results of this
paper. Explicit simulations are performed for Range and two-way
Doppler between Earth and the Cassini spacecraft during its cruise
between Jupiter and Saturn (from May 2001). In this paper, we focus on
two classes of alternative theories: the Post-Einsteinian Gravity and
MOND External Field Effect~\cite{blanchet:2011ys,blanchet:2011zr}.  We
now briefly introduce these two alternative theories of gravity in the
remainder of this section.

\subsection{Post-Einsteinian Gravity (PEG):}
The first alternative metric theory considered is Post-Einsteinian
Gravity (PEG in the following)~\cite{jaekel:2005zr,jaekel:2005vn,jaekel:2006kx,jaekel:2006uq,reynaud:2007fk}.
In this theory, the geometric features of general relativity such
as the identification of gravitational fields with the metric and
the equivalence principle are preserved but the form of the
Einstein equations is modified. This
theory relies on the existence of a quantized gravitation and is non-local because of radiative corrections. The
relation between the curvature and the stress energy tensor (which
is local in GR) is generalized such that it takes the
form of a non local response relation~\cite{jaekel:2006kx}. The Einstein equations for a static spherical body
are characterized by two running constants which take the place of
the Newton gravitation constant~\cite{jaekel:2005zr}. Within the perturbative approximation valid in
the Solar System, the metric is characterized by two potentials
$\phi_N$ and $\phi_P$. In isotropic gauge, the metric tensor for a
spherical source can be written as
\numparts   
\begin{eqnarray}
    g_{00} & = & 1+2\phi_N = 1+2\phi(r) + 2 \phi(r)^2+ 2 \delta \phi_N(r)\label{pegmetric1}\\
    g_{ij} & = & \delta_{ij}\left( -1+2\phi_N-2\phi_P  \right)=\delta_{ij}\left( -1+2\phi(r)+2\delta\phi_N-2\delta\phi_P  \right)\label{pegmetric2}
\end{eqnarray}
\endnumparts           
where $\phi(r)=-\frac{GM}{rc^2}$ is the Newtonian potential with
$G$ the Newton constant, $M$ the mass of the central body, $c$
the speed of light, $r$ the radial coordinate while
$\delta\phi_N(r)$ and $\delta \phi_P(r)$ are two functions of the
position characterizing the deviation from GR. The Post-Newtonian
formalism (PPN) is recovered for particular potentials    
\begin{eqnarray*}
  \delta \phi_N(r)&=&(\beta-1)\phi(r)^2 \\
  \delta \phi_P(r)&=&-(\gamma -1)\phi(r)  +(\beta-1)\phi(r)^2 \cdot
\end{eqnarray*}
In order to derive constraints on these functions, we
will consider a series expansion of the two
potentials~\cite{lamine:2011fk}
\numparts
\begin{eqnarray}
   \delta\phi_N(r) & =   & \alpha_1 r + \alpha_2 r^2 +\frac{GM}{c^2\Lambda}\log\frac{r}{\Lambda}\label{dphin} \\
   \delta\phi_P(r) & =   & \chi_1 r + \chi_2 r^2 - \delta\gamma\frac{GM}{c^2r} \label{dphip}
\end{eqnarray}
\endnumparts
where $\alpha_{1,2}$, $\chi_{1,2}$ and $\Lambda$ are PEG
parameters and $\delta\gamma=\gamma-1$ is the traditional PPN
parameter. These coefficients are related to coefficients
appearing in the generalized Einstein field equations which have
the form of a non-local relation between the curvature and the
stress energy tensor.

The expansion (\ref{dphin}-\ref{dphip}) can also be seen from the perspective
of the Ricci tensor. In vacuum, the GR Ricci tensor vanishes, the
spatial part of the PPN Ricci tensor decreases as $1/r^3$. The
extension to the above metric gives a Ricci tensor with a spatial
part decreasing as $1/r^2$ (for the logarithmic term), as $1/r$
(for the linear term) or remaining constant  (for the quadratic
term). It can be noted that a linear and a
quadratic term in the space-time metric naturally appear in
conformal theory of gravity~\cite{mannheim:1989ly} and in this
context can also be invoked to explain certain galactic
observations requiring dark matter. The logarithmic
term produces a $1/r$ modification of the Newtonian gravitational
force which can be used to explain certain observations requiring
dark matter~\cite{tohline:1983fk,kuhn:1987fk,hehl:2009uq}.

\subsection{MOND External Field Effect (EFE):}
The second alternative theory of gravity considered is the
External Field Effect (EFE in the following) produced by
MOND theory. The MOND theory~\cite{milgrom:1983fk} consists in a
modification of the gravity law at low acceleration. Naively one
would therefore expect no significant modification in the solar
system, where the gravitational acceleration is large.
Nevertheless Milgrom~\cite{milgrom:2009vn} and later Blanchet and
Novak~\cite{blanchet:2011ys} have shown that MOND theory predicts
a violation of the strong equivalence principle which implies that
the dynamics of a system is influenced by an external
gravitational field. This EFE implies the presence of an anomalous
quadrupolar correction in the Newtonian potential
~\cite{blanchet:2011ys,milgrom:2009vn}
\begin{equation}
\label{eq:EFE}
    \phi=-\frac{GM}{r}-\frac{Q_2}{2}x^ix^j\left(e_ie_j-\frac{1}{3}\delta_{ij}\right)
\end{equation}
where $e_i$ is a unitary vector pointing towards the galactic
center. The value of the quadrupole $Q_2$ can be computed from the
theoretical model of MOND and depends on the MOND interpolating
function. Let us mention that in Blanchet and Novak~\cite{blanchet:2011ys}, the value
of $Q_2$ has been determined numerically and is framed by two
limits,
\begin{equation}
    2.1 \times 10^{-27} \ s^{-2}\leq Q_2\leq 4.1 \times 10^{-26} \ s^{-2}
\end{equation}
depending on the MOND function used. The modification of the
metric deriving from this modification of the Newtonian potential
can be expressed using the metric parametrization
(\ref{pegmetric1}-\ref{pegmetric2}) with
\numparts
\begin{eqnarray}
    \delta \phi_N & = &  -\frac{Q_2}{2c^2}x^ix^j\left(e_ie_j-\frac{1}{3}\delta_{ij}\right)\label{dphimond}\\
\delta \phi_P & = & 0.
\end{eqnarray}
\endnumparts
Blanchet and Novak~\cite{blanchet:2011ys,blanchet:2011zr} have
also shown that this quadrupolar term implies the existence of a
secular precession of planetary perihelia. New
improved INPOP results~\cite{fienga:2011qf} on planetary
perihelion precession put an even more stringent constraint on the
quadrupole~\cite{blanchet:2011zr}.

\section{Numerical simulations of observables from the space-time metric}\label{sec:num}
In this section, we present the numerical methods used to simulate
Range and Doppler signals directly from the space-time metric.
First of all covariant definitions of the observables are given.
After, we explain in detail the different steps needed to simulate
signals from the metric. This includes the derivation and
integration of the equations of motion, the derivation and
integration of the equation of proper-time and the computation of
the propagation of light in curved space-time and the
determination of the observables.

\subsection{Tracking observables} \label{sec:observables}

Observables, that is to say measured quantities, are by definition
gauge invariant~: they do not depend on the choice of a coordinate system. The
simulations, for example the integration of the equations of
motion, are necessarily done in a particular coordinate system
(different equations of motion representing the same situation in
different coordinates system can be found
in~\cite{soffel:1989fk,brumberg:1991uq}). A reduction of
coordinates, that is to say a transformation of
coordinate-dependent quantities to measurable quantities
(observables) is done in the software and presented below.

The situation corresponding to traditional radioscience
measurements is the following~: an electromagnetic signal is sent
by an emitter (denoted by subscript $e$), eventually re-transmitted by
a transponder (denoted by $t$) and received by an observer (denoted by $r$)
which is often the same as the emitter. The emitted signal is
characterized by its proper frequency $\nu_e$ and by the emission
proper time $\tau_e$ (time when the signal is sent as given by an
ideal clock moving with the emitter). Similarly the received
signal is characterized by its proper frequency $\nu_r$ and by the
reception proper time $\tau_r$.

The Range signal (evaluated at reception) is related to the
signal propagation time between the emitter and the receiver:
\begin{equation}
    R(\tau_r)=c(\tau_r-\tau_e) \cdot
\end{equation}
The Doppler signal is related to the frequency shift between the emitter and the receiver:
\begin{equation}
    D(\tau_r)=\frac{\nu_r}{\nu_e}  \cdot
\end{equation}

These definitions are based on proper quantities that are
measurable.

\subsection{Equations of motion}
The equations of motion are directly derived from the metric
using the geodesic
equations~\cite{misner:1973fk,brumberg:1991uq}, integrated with
respect to coordinate time $t$:
\begin{equation}
    \frac{1}{c^2}\frac{d^2x^i}{dt^2}=-\Gamma^i_{00}-2\Gamma^i_{0j}
\beta^j-\Gamma^i_{jk}\beta^j \beta^k + \Gamma^0_{00}\beta^i+2\Gamma^0_{0j}\beta^i \beta^j+\Gamma^0_{jk}
\beta^i \beta^j \beta^k \label{eq:eom}
\end{equation}
\noindent where $x^i$ are the spatial coordinates of the test
particle, $\beta^i=v^i/c$ is the reduced coordinate
velocity, $\Gamma^\alpha_{\beta\gamma}$ are the Christoffel
symbols of the considered metric (Greek indices run from $0$ to
$3$ while Latin indices from $1$ to $3$) and $t$ is coordinate
time. The Christoffel symbols are computed using the metric and
its first derivatives
\begin{equation}
  \Gamma^\alpha_{\beta\gamma}=\frac{1}{2}g^{\alpha\delta}
\left(g_{\delta\gamma,\beta}+g_{\beta\delta,\gamma}-g_{\beta\gamma,\delta}\right)
\end{equation}
with $g^{\alpha\delta}$ the inverse of the metric
$g_{\alpha\delta}$,
$g^{\alpha\delta}g_{\delta\beta}=\delta^\alpha_\beta$. In our
approach, we choose to implement analytically the derivative of
the metric so that the Christoffel symbols and the right hand side
of \Eref{eq:eom} can be computed exactly. Let us mention
the other possibility consisting in implementing a numerical
derivative of the space-time metric~\cite{hees:2010fk}.
Nevertheless, the numerical accuracy of the derivative can be
problematic and is time-consuming.

The software is independent of any ephemerides. This choice is
justified since external ephemerides (such as
INPOP~\cite{fienga:2009kx,fienga:2011qf}, DE~\cite{newhall:1983uq} or
EPM~\cite{pitjeva:2005kx}) are computed in General Relativity or
within PPN
formalism~\cite{fienga:2010vn,folkner:2010zr,pitjeva:2010ys} and the
goal of our approach is to go beyond the latter. In order to be fully
consistent, we produce the ephemerides needed by integrating the
equations of motion of all bodies considered in the problem, here the
spacecraft, the Sun and the Earth (observer), in the theory considered.

\subsection{Clock behavior} \label{sec:clock}
In the process of reduction to relativistic observables, the
proper time equation is integrated for each body considered, and
in particular for the clocks used in the Doppler/Range
measurements. This equation is
\begin{equation}
\label{dtau}
 \frac{d\tau}{dt}=\sqrt{g_{00}+2g_{0i}\beta^i+g_{ij}\beta^i\beta^j}
\end{equation}
where $g_{\mu\nu}$ is the space-time metric and $\tau$ the proper
time. This integration is performed on the trajectory of the clock
(this trajectory has been computed before, see previous section).
As a result of this integration, we get the relation between
proper time and coordinate time for the different clocks
$\tau_i(t)$.

\subsection{Light propagation}
Finally, the signal propagation in the gravitational field has to
be modelled. This is done thanks to the Synge World function
formalism and the use of the time transfer
function~\cite{le-poncin-lafitte:2004cr,teyssandier:2008nx}.
Within this formalism, the time transfer (and the frequency shift)
can be expressed as an integral of some function defined from the
metric (and its derivatives) along the Minkowski path of the
photon. From a theoretical point of view, this method is
equivalent to finding the solution of the null geodesic but from a
practical point of view, this method avoids the explicit
resolution of the null geodesic. More precisely, instead of
solving the null geodesic (which is a boundary value problem), we
can integrate some functions defined by the metric and its
derivatives over the Minkowski path of the photon (the form of the
function to integrate is given to all orders
in~\cite{teyssandier:2008nx}). In this section, we briefly
describe how to compute the coordinate propagation time.

Following~\cite{teyssandier:2008nx}, the reception time transfer function $\mathcal T_r$ is defined by
\begin{equation} \label{timefun}
    t_r-t_e=\mathcal T_r(\bm x_e,t_r,\bm x_r)
\end{equation}
where $t_r$ and $t_e$ are coordinate times related to the
reception and the emission of the signal, $\bm x_e$ and $\bm x_r$
are the coordinate positions of the emitter (at emission time) and
of the receiver (at reception time). The expression of the time
transfer function is given in~\cite{teyssandier:2008nx}:
\begin{equation}    \label{tec}
      \mathcal T_r(\bm x_e,t_r,\bm x_r) =\frac{1}{c}R_{er}+\frac{1}{c}\Delta_r (\bm x_e,t_r,\bm x_r)
\end{equation}
with $R_{er}$ the Euclidean distance between the emission and
reception points $R_{er}=\left| \bm  x_r(t_r)  - \bm
x_e(t_e)\right|$ and $\Delta_r$ the gravitational contribution to
the time transfer, that is the traditional Shapiro time
delay. In the case of a moving source, the last equation is
implicit since the position of the emitter (at emission time)
depends on the time transfer function: $\bm x_e(t_e)=\bm
x_e(t_r-\mathcal T_r) $. In the following, the determination of
the time transfer is performed up to order $1/c^3$. This is
sufficient for most of current space missions but the same
approach can be implemented to higher order. The software proceeds
in two steps: first, it computes the Minkowskian emission time
$t_{em}$ and then the gravitational time delay $\Delta_r$.

The first step is the determination of $t_{em}$, the Minkowskian emission time computed in flat space-time, which is solution of
\begin{equation} \label{tem}
    t_r-t_{em}=\frac{\left| \bm x_e(t_{em}) - \bm x_r(t_r) \right|}{c}\cdot
\end{equation}

The last equation is implicit and can be solved numerically by iteration. The iterative procedure is standard:
\begin{eqnarray}
  {Start:}   &\qquad& t_{em}^{(0)}=t_r-\frac{\left| \bm
      x_e(t_{r}) - \bm x_r(t_r) \right|}{c} \label{itstart}\\
  {Loop:}   &\qquad&t_{em}^{(i+1)}=t_r-\frac{\left| \bm
x_e(t_{em}^{(i)}) - \bm x_r(t_r) \right|}{c}\\
  {End:} &\qquad& {when}\;\;\left| t_{em}^{(i+1)} -
t_{em}^{(i)}\right|<\epsilon                    \label{itend}
\end{eqnarray}
with $\epsilon$ being the desired accuracy. In practice, this
procedure is very efficient and converges in two or three
iterations. An alternative method to determine the Minkowskian
emission time consists in expanding \Eref{tem} up to order
$1/c^3$~\cite{petit:1994uq}. This gives
\begin{eqnarray}
    t_{em}&=&t_r-\frac{D_{er}}{c}-\frac{\bm D_{er}\cdot \bm v_e(t_r)}{c^2}\nonumber\\
    &&-\frac{D_{er}}{2c^3}\left[v_e(t_r)^2+\left(\frac{\bm D_{er}\cdot\bm v_e(t_r)}{D_{er}}\right)^2-\bm a_e(t_r)\cdot \bm D_{er}\right]+\mathcal O(1/c^4)
\end{eqnarray}
where $\bm D_{er}=\bm x_r(t_r)-\bm x_e(t_r)$ and $D_{er}=\left|\bm
D_{er}\right|$, $\bm v_e$ is the emitter velocity at reception and
$\bm a_e$ its acceleration.  We checked that the two methods give
the same results. Nevertheless, the iterative method is more
precise (and also valid to higher order).

The second step is the computation of the gravitational time delay
$\Delta_r/c$. To this aim, we introduce a post-Minkowskian
decomposition of the metric $h_{\mu\nu}=g_{\mu\nu}-\eta_{\mu\nu}$
with $\eta_{\mu\nu}$ the Minkowski metric. The post-Minkowskian
metric $h_{\mu\nu}$ is of order $G/c^2$ where $G$ is the
gravitational constant. With these definitions, the gravitational
correction of \Eref{tec} to first post-Minkowkian order (that
is to say to order $G/c^2$) is given by~\cite{teyssandier:2008nx}
\begin{equation} \label{deltar}
\Delta_r (\bm x_e,t_r,\bm x_r)  =\frac{R_{er}}{2}\int_0^1 f(z^\alpha(\mu)) d\mu
\end{equation}
where $R_{er}=\left| \bm x_r(t_r)-\bm x_e(t_e)\right|$,
\begin{equation} \label{fmet}
    f(z^\alpha)=-h_{00}-2N^i_{er}h_{0i}-N^i_{er}N^j_{er}h_{ij}
\end{equation}
and the integration path $z^\alpha(\mu)$ is the Euclidean straight line between the emitter and the receiver
\numparts
\begin{eqnarray}
z^0(\mu)     & =  &ct_r-\mu R_{er} \label{z0} \\
z^i(\mu) &=&   x^i_r(t_r)-\mu R_{er} N^i_{er}.\label{zi}
\end{eqnarray}
\endnumparts
The unit vector $N^i_{er}$ points from the emitter to the receiver
\begin{equation}
    \bm{N}_{er}=\frac{\bm{x}_r(t_r)-\bm{x}_e(t_e)}{R_{er}}\cdot
\end{equation}
Let us recall that the previous formulas can be extended to higher
order if necessary.

From \Eref{fmet}, one sees that $\Delta_r$ is of order
$1/c^2$. Writing $t_e=t_{em}+\delta t_e$, \Eref{tec} gives
\begin{equation}
    t_r-t_{em}-\delta t_e = \frac{\left|\bm x_r(t_r)-\bm x_e(t_{em} + \delta t_e) \right|}{c}+\frac{1}{c}\Delta_r (\bm x_e(t_{em}+\delta t_e),t_r,\bm x_r(t_r))
\end{equation}

Since $\Delta_r/c$ is already of order $1/c^3$, we can drop the $\delta t_e$ term in $\Delta_r$. After expanding the first term, we get
\begin{equation*}
    \delta t_e=-\frac{1}{c}\Delta_r (\bm x_e(t_{em}),t_r,\bm
        x_r(t_r)) \left( 1-\frac{\bm v_e(t_{em})\cdot\bm N_{er}}{c}\right)^{-1}
\end{equation*}

Up to order $1/c^3$, we finally obtain
\begin{equation}         \label{te}
    t_e=t_{em}-\frac{1}{c}  \Delta_r (\bm x_e(t_{em}),t_r,\bm x_r(t_r)) +\mathcal O(c^{-4})
\end{equation}
where $t_{em}$ is computed iteratively by
(\ref{itstart}-\ref{itend}) and $\Delta_r$ is determined by the
integral~(\ref{deltar}) performed on the Euclidean path between
the emitter and the receiver.

As an example, if the metric used is the Schwarzschild metric in
isotropic coordinates
$ds^2=(1-2\frac{m}{r})c^2dt^2-(1+2\gamma\frac{m}{r})d\bm x^2$
(with $m=GM/c^2$), the integration of $\Delta_r$ gives the usual
logarithmic term in the Shapiro delay
\begin{equation} \label{shapiro}
    \Delta_r=(1+\gamma) m\ \ln \left[\frac{r_e+r_r+R_{er}}{r_e+r_r-R_{er}} \right]\cdot
\end{equation}

The method based on the time transfer function $\mathcal{T}_r$ is
very efficient and easy to implement numerically. This method as
presented above is valid only to $1/c^3$ but can be generalized to
higher order (see \cite{teyssandier:2008nx}). Moreover, this
method avoids the explicit resolution of the null geodesic in
curved space-time. The computation of the null geodesic is more
delicate since it is a boundary value problem (BVP) that needs to
be solved by a shooting method~\cite{san-miguel:2007hc}.

\subsection{Range observable}
From the coordinate propagation time determined in the last section, it is straightforward to determine the Range observable as a function of the reception proper time
\begin{equation}
    R(\tau_r)=c(\tau_r-\tau_e(\tau_r))\cdot
\end{equation}

The determination of $\tau_e$ from $\tau_r$ is done in three steps:
\begin{enumerate}
    \item conversion from the reception proper time $\tau_r$ to the reception coordinate time $t_r$;
    \item  computation of the coordinate time transfer using~(\ref{te});
    \item  transformation from the coordinate emission time $t_e$ to the emission proper time $\tau_e$.
\end{enumerate}

The transformation from proper time $\tau_r$ to coordinate
time $t_r$ is done by inverting numerically the relation $\tau(t)$
given by the integration of (\ref{dtau}). This inversion is done
by a Newton method. The transformation from coordinate time
$t_e$ to proper time $\tau_e$ is simply done by the evaluation of
the relation $\tau(t)$.

\subsection{Doppler observable}
The Doppler is modeled as the ratio between the received signal
frequency and the emitted signal frequency.
Following~\cite{blanchet:2001ud}, we can write the Doppler signal
as
\begin{equation}\label{dop1}
    D(\tau_r)=\frac{\nu_r}{\nu_e}=\frac{d\tau_e}{d\tau_r}=\left(\frac{d\tau}{dt}\right)_e\frac{dt_e}{dt_r}\left(\frac{dt}{d\tau}\right)_r
\end{equation}
where $d\tau_{e/r}$ represents the proper period of the emitted/received photon.

The first and the last factor of the previous equation are easily
determined from the metric through \Eref{dtau}. The second
term of (\ref{dop1}) is more difficult. From (\ref{timefun}), we
can write $t_e=t_r-\mathcal T_r(\bm x_e,t_r,\bm x_r)$. The
derivative of this relation gives
\begin{eqnarray*}
    \frac{dt_e}{dt_r}&=&1-\frac{d\mathcal T_r(\bm x_e,t_r,\bm x_r)}{dt_r}\\
    &=&1-\frac{\partial \mathcal T_r}{\partial \bm x_e}\cdot \bm v_e(t_e)\frac{dt_e}{dt_r}-\frac{\partial\mathcal T_r}{\partial t_r}-\frac{\partial\mathcal T_r}{\partial \bm x_r}\cdot \bm v_r(t_r) .
\end{eqnarray*}
This gives
\begin{equation} \label{dtedtr}
    \frac{dt_e}{dt_r}=\frac{1-\frac{\partial\mathcal T_r}{\partial t_r}-\frac{\partial\mathcal T_r}{\partial \bm x_r}\cdot \bm v_r(t_r)}{1+\frac{\partial \mathcal T_r}{\partial \bm x_e}\cdot \bm v_e(t_e)}\cdot
\end{equation}

This expression was already derived in~\cite{jaekel:2006uq} and is
consistent with Eq.~(A.46) of Blanchet et al.~\cite{blanchet:2001ud}
\begin{equation}
    \frac{dt_e}{dt_r}=\frac{(k_0)_r}{(k_0)_e}\frac{1+\left(\frac{k_i}{k_0}\right)_r\frac{v^i_r}{c}}{1+\left(\frac{k_i}{k_0}\right)_e\frac{v^i_e}{c}}
\end{equation}
where $k^\mu$ is the photon wave vector, and with the expressions
of the photon wave vector given in Teyssandier and Le
Poncin-Lafitte (relations (40-42) from~\cite{teyssandier:2008nx})
\begin{eqnarray}
\left(\frac{k_i}{k_0}\right)_r   & =  &-c\frac{\partial \mathcal T_r}{\partial x^i_r}\left[1-\frac{\partial \mathcal T_r}{\partial t_r}\right]^{-1}  \\
\left(\frac{k_i}{k_0}\right)_e   & =  &c\frac{\partial \mathcal T_r}{\partial x^i_e} \\
\frac{(k_0)_r}{(k_0)_e}  & =  & 1-\frac{\partial \mathcal T_r}{\partial t_r}.
\end{eqnarray}
The last four equations are equivalent to~(\ref{dtedtr}). Finally, the quantities $\partial \mathcal T_r/\partial \bm x_{e/r}$ and $\partial \mathcal T_r/\partial t_{r}$ are obtained from~(\ref{tec})
\begin{eqnarray}
c   \frac{\partial \mathcal T_r}{\partial \bm x_{e/r}}&=&\mp \bm N_{er}\mp \frac{\bm N_{er}}{R_{er}}\Delta_r  \label{dTdx}\\
&+&\frac{R_{er}}{2}\int_0^1\left[\frac{\partial f}{\partial z^\alpha}\frac{\partial z^\alpha}{\partial \bm x_{e/r}}+\frac{\partial f}{\partial N^i_{er}}\frac{\partial N^i_{er}}{\partial \bm x_{e/r}}\right]d\mu \nonumber \\
    \frac{\partial \mathcal T_r}{\partial t_r}&=&\frac{R_{er}}{2}\int_0^1\frac{\partial f}{\partial z^0}d\mu  \label{dTdt}
\end{eqnarray}
where the integrals are performed over the Euclidean straight line
between emitter and receiver as parameterized
by~(\ref{z0}-\ref{zi}). The derivatives appearing in the integrand
can easily be expressed using the expression~(\ref{fmet}) of the
function $f$ and~(\ref{z0}-\ref{zi})
\numparts
\begin{eqnarray}
    \frac{\partial f}{\partial z^\alpha} &=  & -h_{00,\alpha}-2h_{0i,\alpha}N^i_{er} -h_{ij,\alpha}N^i_{er}N^j_{er}\label{dcalc1}\\
    \frac{\partial f}{\partial N^i_{er}}&=& -2h_{0i}-2h_{ij}N^j_{er}\\
    \frac{\partial z^0}{\partial x^i_{e/r}}&=&\pm \mu N^i_{er}\\
    \frac{\partial z^j}{\partial x^i_e}&=&\mu\delta_i^j\\
        \frac{\partial z^j}{\partial x^i_r}&=&(1-\mu)\delta_i^j\\
        \frac{\partial N^j_{er}}{\partial x^i_{e/r}}&=&\mp\frac{\delta_i^j-N^i_{er}N^j_{er}}{R_{er}}  \label{dcalc2} \cdot
\end{eqnarray}
\endnumparts

In summary, the observable frequency shift can be computed from
\begin{equation}
    D(\tau_r)=\frac{\left[\sqrt{g_{00}+2g_{0i}v^i+g_{ij}v^iv^j}\right]_{x_{e}}}{\left[\sqrt{g_{00}+2g_{0i}v^i+g_{ij}v^iv^j}\right]_{x_{r}}} \times \quad \frac{1-\frac{\partial\mathcal T_r}{\partial t_r}-\frac{\partial\mathcal T_r}{\partial \bm x_r}\cdot \bm v_r(t_r)}{1+\frac{\partial \mathcal T_r}{\partial \bm x_e}\cdot \bm v_e(t_e)} \label{dshapiro}
\end{equation}
 where the derivatives of the time transfer function are computed with integrals~(\ref{dTdx}-\ref{dTdt}) and the relations (\ref{dcalc1}-\ref{dcalc2}).

 As an example, with the Schwarzschild metric  $ds^2=(1-2\frac{m}{r})c^2dt^2-(1+2\gamma\frac{m}{r})d\bm x^2$, the computation of the integrals ~(\ref{dTdx}-\ref{dTdt}) gives  
\numparts
\begin{eqnarray}
  \frac{dt_e}{dt_r}& =  & \frac{q_r}{q_e} \label{dtedtr_1}\\
q_e&=&1-\frac{\bm N_{er}\cdot \bm v_e}{c}-\frac{2(1+\gamma)GM}{c^3}\frac{\bm N_{er}\cdot \bm v_e(r_e+r_r)+R_{er}\frac{\bm x_e\cdot \bm v_e}{r_e}}{(r_e+r_r)^2-R_{er}^2}\\
q_r&=&1-\frac{\bm N_{er}\cdot \bm v_r}{c}-\frac{2(1+\gamma)GM}{c^3}\frac{\bm N_{er}\cdot \bm v_r(r_e+r_r)-R_{er}\frac{\bm x_r\cdot \bm v_r}{r_r}}{(r_e+r_r)^2-R_{er}^2}\label{dtedtr_3}
\end{eqnarray}     
\endnumparts
which is exactly equivalent to the results of Blanchet et
al~\cite{blanchet:2001ud}. But we recall that simply using
(\ref{dtedtr_1})-(\ref{dtedtr_3}) is not sufficient for our case,
as we want to keep a formulation which remains valid for any metric, hence the use of the general formulation
(\ref{dshapiro}).

The Doppler is evaluated by the integrating functions defined
from the metric and its first derivative over the Euclidean straight
line between the emitter and the receiver. Once again, this method
avoids the computation of the null geodesic in curved space-time
and can be extended to higher order if necessary.
\\

The implementation of the different steps presented above allows
us to simulate radioscience observables directly from the
space-time metric. With this approach, it is easy to change the
underlying gravitation theory, and therefore produce signals that
would be observed in general relativity and alternative theories
of gravity.

\section{Comparison between signals produced in different
  theories} \label{sec:comp} The previous sections presented how to
simulate Range and Doppler signals of different space missions in
general relativity and in alternative theories of gravity. In this
section, we will describe the method used to compare the signal in an
alternative theory of gravity with the signal in General Relativity. A
direct comparison does not provide useful information.  Indeed, even if the Range and Doppler are observables (i.e. measured quantities) that are independent of any coordinate system (see \sref{sec:observables}), the simulations depend on the initial conditions of the different bodies (Earth and spacecraft) which are coordinate dependent. The procedure to extract the influence of the initial conditions on the observables consists of performing a fit of the initial conditions of the spacecraft, the Earth, and the mass of the Sun (in fact the product $GM$).  
More precisely, we treat signals simulated in
an alternative theory of gravity as ``real'' observations and we
analyze them in GR. This analysis consists in a least-squares fit of
the different parameters. The residuals of this fit then display an incompressible deterministic signature directly related to the modification of the considered alternative gravitation theory independently of any coordinate system. In a next step (cf. Introduction) this template can be systematically searched for in the residuals from real data.

\subsection{Least-squares fit}

As mentioned, the Range/Doppler signals generated in an alternative
theory of gravity will be analyzed in GR with a fit of the initial
conditions and of the mass of the Sun.  This fit consists in
minimizing the quantity
\begin{equation}
    S=\sum_i \frac{(R_s(\tau_i)-R_{GR}(\tau_i,p_l))^2}{\sigma_{Ri}^2}  + \sum_i \frac{(D_s(\tau_i)-D_{GR}(\tau_i,p_l))^2}{\sigma_{Di}^2}
\end{equation}
where $R_s(\tau_i)$ and $D_s(\tau_i)$ are the simulated Range and
Doppler in an alternative theory of gravity at observation time
$\tau_i$, $R_{GR}(\tau_i,p_l)$ and $D_{GR}(\tau_i,p_l)$ are the Range
and Doppler (at observation time $\tau_i$) simulated in General
Relativity with the different parameters $p_l$ ($p_l$ represents the
different parameters to be fitted, i.e. the initial conditions and the
masses of the planets) and $\sigma^2_{Ri/Di}$ are the Range/Doppler
variances at time $\tau_i$. Would the fit be performed with real data,
these variances would correspond to the accuracy of the
measurements. In our case, since we work with simulations and not real
data, we will assume constant uncertainties $\sigma_{Ri}=\sigma_R$ and
$\sigma_{Di}=\sigma_{D}$ corresponding to Range and Doppler
accuracies of the considered mission.

The scenario to perform the fit is standard and can be found
in~\cite{zarrouati:1987fk,peters:1981il,lainey:2004uq,lainey:2004tw}.
The fit is produced by an iterative procedure. At each iteration,
the quantity to minimize $S$ is linearized with respect to the
parameters $p_l$. Denoting by $O$ either one of the observables $R$
or $D$ and denoting by $\sigma_i$ either $\sigma_{Ri}$ or
$\sigma_{Di}$, we can write
\begin{eqnarray}
    S&=&\sum_i \frac{(O_s(\tau_i)-O_{GR}(\tau_i,p_l))^2}{\sigma_i^2}\\
     &=& \sum_i \frac{\left( O_s(\tau_i)-O_{GR}(\tau_i,p_l^{(0)})
             - \frac{\partial O_{GR}(\tau_i,p_l^{(0)}) }{\partial p_j}\delta p_j\right)^2}{\sigma_i^2}    \nonumber
             \end{eqnarray}

The variation of the parameters minimizing this quantity is given by
\begin{equation}       \label{leastsquare}
  {  \delta p_l}=( B^T  B)^{-1} B^T\left( O_s  - O_{GR}(p_l^{(0)} )   \right)
\end{equation}
where $B$ is the matrix of the partial derivatives
\begin{equation}
    B_{ij}=\frac{\partial O_{GR}}{\partial p_j}(\tau_i,p_l^{(0)}),
\end{equation}
$O_s$ is the vector containing the simulated observations
($O_s(\tau_i)$)  and $O_{GR}(p_l^{(0)} )$ is the vector containing
the values simulated in General Relativity with the initial
conditions $p_l^{(0)}$ ($O_{GR}(\tau_i,p_l^{(0)})$).

The analysis of a signal simulated in an alternative theory of gravity
$O_s$ consists in iterating the least-squares fit
(\ref{leastsquare}).

\subsection{Simulations of the observables and of the partial derivatives in General Relativity}

The least-squares fit needed in order to compare signals in
different theories involves the computation of the observable in
GR ($O_{GR}$ in (\ref{leastsquare})). This simulation can be done
using the software developed in \sref{sec:num} with GR
space-time metric. Nevertheless, since the derivative of the
observable is also needed, we develop analytically the equations
used for the fit and we compute analytically the partial
derivatives of these equations.

The equations of motion in GR are the Einstein-Infeld-Hoffman (EIH)
equations obtained from the 1PN metric (in harmonic coordinates) for
point masses~\cite{misner:1973fk,soffel:1989fk}. The derivatives of
these equations of motion with respect to the parameters involved in
the fit give the variational equations, which are obtained after
lengthy but straightforward calculations.

The equation of proper time is also obtained from the 1PN metric
in harmonic coordinates and is given by the IAU 2000
resolutions~\cite{soffel:2003bd}. The variational equations of the
proper time have also been computed analytically.

The integration of the equations of motion, of the equation of proper time and of the variational equations are performed numerically. 

Finally, the computation of the light propagation has been done
with the same spacetime metric. The resulting formulas can be
found in the
literature~\cite{blanchet:2001ud,le-poncin-lafitte:2004cr} and are
also given in Equations~(\ref{shapiro}) and (\ref{dtedtr_1}-\ref{dtedtr_3}). The
partial derivatives of these expressions have also been computed
analytically.

These computations provide the quantity $O_{GR}$ and its
partial derivatives which are needed for the fit.

\subsection{What can be fitted ?}

In the previous section, we showed that the comparison of signals
coming from different theories of gravity requires a least-squares
fit of the different parameters involved in the problem. The main
parameters are the initial conditions of the bodies (planets and
spacecraft) and the Sun $GM$. One has to be careful when fitting the initial
conditions of the bodies because of correlations between the
parameters. In some cases, these correlations can become
theoretically equal to one. As a consequence, the matrix $B^TB$
becomes degenerate and is not invertible which poses difficulties
for the solution of the least-squares problem~(\ref{leastsquare}). This problem of rank deficiency is more general and is associated with symmetries~\cite{bonanno:2002fk}. In the case considered in this paper (simulations of Cassini spacecraft, the Earth and the Sun), the Range/Doppler signals are invariant under global translations and rotations. Therefore, we have a rank deficiency of order 9. In practice, we fit the 6 initial conditions of Cassini spacecraft, 3 initial conditions of the Earth and the Sun $GM$. The 9 others initial conditions (6 for the Sun and 3 for the Earth) are fixed to avoid any degeneracy and correspond to fixing the origin and the orientation of the axes.

\subsection{Numerical accuracy}
\begin{figure}[htb]
\begin{center}
\includegraphics[width=0.7\textwidth]{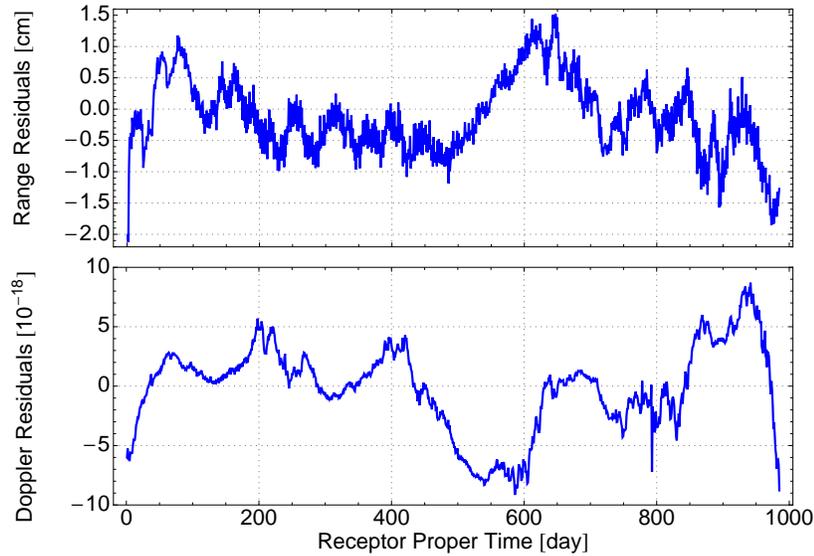}
\end{center}
\caption{Representation of the Range (top) and Doppler (bottom)
  residuals for a simulation done in GR followed by a least-squares
  fit of the initial conditions using a GR model. These residuals are only due to
  numerical error. The orders of magnitude of the signals are $10^9$~km
  for the Range and $10^{-4}$ for the Doppler which means that the
  relative accuracy of the software is around
  $10^{-13}-10^{-14}$. This simulation has been performed for a
  two-way link to the Cassini spacecraft starting from May 2001.}
\label{figPrecision}
\end{figure}

A consistency test has been performed in order to check the numerical
accuracy of our simulations. The test consists in a simulation of
observations in General Relativity with the software presented in \sref{sec:num} followed by an analysis of these simulations with
the least-squares fit in GR after changing the initial conditions. The
obtained residuals are only due to numerical
errors. \Fref{figPrecision} represents the residuals obtained for
a simulation of a two-way Range/Doppler link from Earth to a spacecraft during 3 years. The initial conditions in the simulation were chosen to be those of Cassini on 1st May 2001. 
 The Range accuracy is of the
order of a centimeter while the Doppler accuracy is of the order of
$10^{-17}$ (in terms of velocity this corresponds to about a nanometer
per second). To have an idea of the relative uncertainty, the Range
signal is of the order of $10^9$~km and the Doppler signal is of the
order of $10^{-4}$ which means that the relative accuracy of the
simulation and of the fit is around $10^{-13}-10^{-14}$. This relative
accuracy corresponds to the expected accuracy of the numerical
integrator.  In our software, two integrators are implemented and can
be used: a Rung-Kutta 45 and a Radau integrator. Finally, two
independent programs have been built and compared with the same level
of accuracy.

%
%
%
%
\section{Simulations in alternative theories of gravity}~\label{sec:results}

In this section, we present results obtained with our software (see
also~\cite{hees:2011vn,hees:2011kx}). The situation considered is the
Cassini 3-years cruise from Jupiter to Saturn. We take the planetary
initial conditions from ephemerides and the Cassini initial conditions from the SPICE and simulate a
Range and a two-way Doppler link between Earth and Cassini
spacecraft. We do not use any real data coming from the Cassini
mission. Instead, we produce data in alternative theories of gravity
and analyze them in GR in order to compare the expected deterministic signature in the (simulated) residuals with the
Cassini precision. The simplified situation considered is the following: the Sun,                               
the Earth and the Cassini spacecraft. We will show below that the addition of another planet (Jupiter for example) does not change significantly the results. 
  
The correct analysis in GR is to fit initial conditions (eg. Earth and spacecraft positions/velocities, sun mass, etc\dots) to the simulated data, then the residuals of the fit (green continuous lines on figures~\ref{figCass_gamma},~\ref{figCass_chi1} and~\ref{figCass_mond}) are the incompressible physically observable signals expected in the residuals of the real data if the theory of gravity in nature is not GR (eg. PEG with $\delta\gamma=10^{-5}$ in \fref{figCass_gamma}, PEG with $\chi_1=10^{-21}\ m^{-1}$ in \fref{figCass_chi1} or with an External Field Effect in \fref{figCass_mond}). An alternative could be to not fit initial conditions in the GR analysis, but use the same initial conditions as in the simulation. The result then represents the difference in observables when starting from the same initial conditions but using different theories, and can give rise to signatures which are significantly larger than when fitting initial conditions (illustrated by the blue dash-dot curves on figures~\ref{figCass_gamma},~\ref{figCass_chi1} and~\ref{figCass_mond}), as the fit of initial conditions always absorbs some of the signal. We show these curves for illustration of that difference, but stress that drawing conclusions from these curves as to the expected signatures in the residuals of a real GR data analysis is incorrect, as in reality it is always necessary to fit initial conditions (they are unknown a priori).

Moreover, we only focus on two
alternative theories of gravity presented in \sref{sec:outline}:
PEG theory and MOND EFE.  In each of the simulations presented below,
only one of the PEG and MOND EFE parameters is not vanishing. A more
general study considering variations of several parameters is postponed
to future work.

\subsection{Simulations in Post-Einsteinian Gravity (PEG)} \label{sec:simuPEG}
\begin{figure}[htb]
\begin{center}
\includegraphics[width=0.7\textwidth]{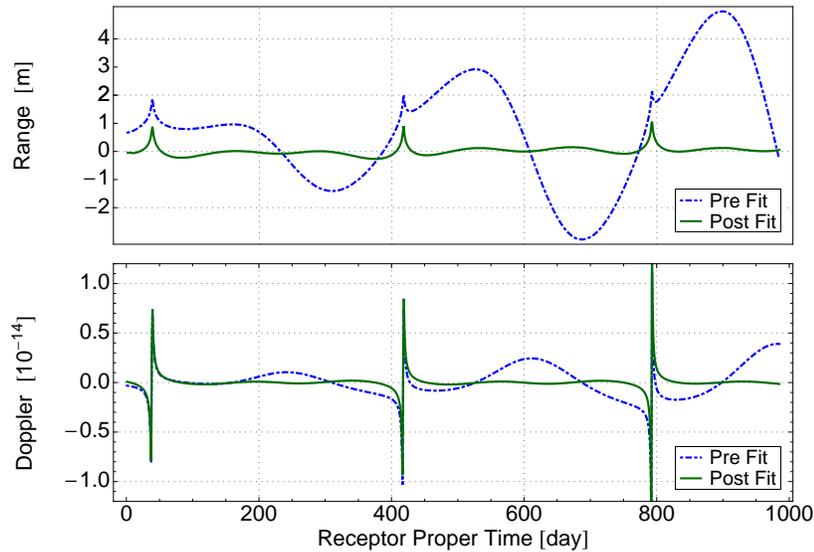}
\end{center}
\caption{Representation of the Range (top) and Doppler (bottom) signals
  due to an alternative theory with $\gamma-1=10^{-5}$ (and all other PEG parameters vanishing). The blue
  (dash-dot) line is the difference between a simulation in
  the alternative theory and a simulation in GR (with the same
  initial conditions) and is for illustration purposes only (see text at the beginning of \sref{sec:results}). The green (continuous) line is the residuals obtained
  after analyzing the simulated data in GR (which means after the fit
  of the different initial conditions using a GR model) and represents the expected physical signal.}
\label{figCass_gamma}
\end{figure}

We use the metric (\ref{pegmetric1}-\ref{pegmetric2}) with the expansion of the
potentials (\ref{dphin}-\ref{dphip}) to determine effects due to PEG on the
Cassini signals. As can be seen from the expression of the metric (\ref{pegmetric1}-\ref{pegmetric2}), only corrections coming from the central body (the Sun) are considered, because these provide the most dominant contribution in the signature of the residuals. Different simulations were performed with
different values of the PEG parameters and then analyzed in GR by
fitting the initial conditions of the Earth, the initial
conditions of Cassini spacecraft and the Sun mass. For example,
\fref{figCass_gamma} represents the Range and Doppler
differences between a simulation in a theory with
$\delta\gamma=\gamma-1=10^{-5}$ (and all other PEG parameters
vanishing) and in GR. The three peaks occur during solar
conjunctions. The blue (dash-dot) curves are the direct
differences of signals generated in PEG theory with signals
generated in GR. 
On the other hand, the green (continuous) lines represent the residuals obtained from simulated data in the alternative theory after
the fit of the initial conditions. These signals are the ones
expected to be detected in the residuals of the analysis of the real
data if the theory of gravity is PEG theory (with
$\delta\gamma=10^{-5}$) and if the analysis is performed with
traditional GR theory. 
\\

\begin{figure}[bt]
\begin{center}
\includegraphics[width=0.7\textwidth]{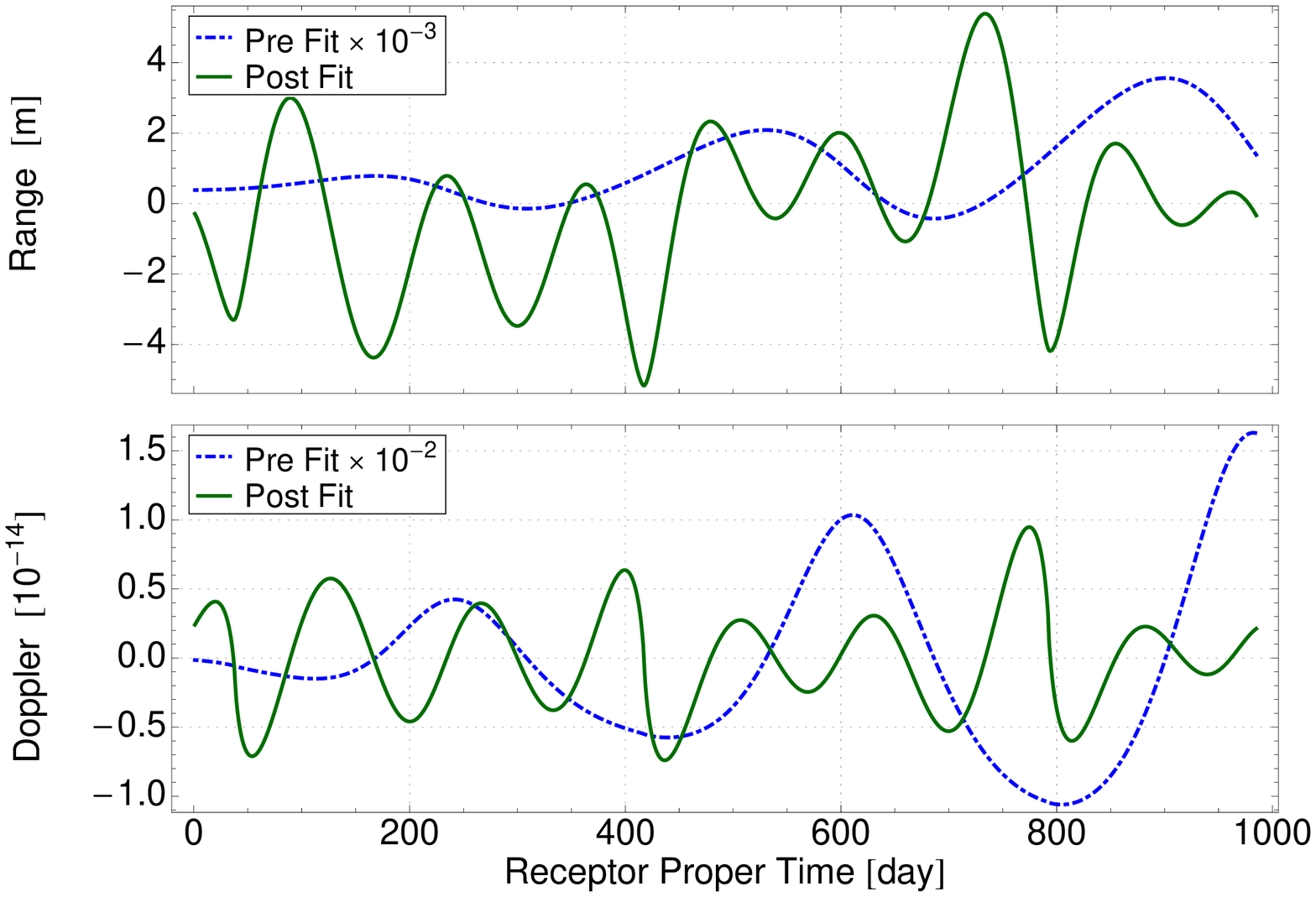}
\end{center}
\caption{Representation of the Range (top) and Doppler (bottom) signals
  due to an alternative theory with $\chi_1= 10^{-21}\ m^{-1}$ (and all other PEG parameters vanishing). The
  blue (dash-dot) line is the difference between a simulation in
  the alternative theory and a simulation in GR (with the same
  initial conditions) and is for illustration purposes only (see text at the beginning of \sref{sec:results}). The green (continuous) line is the residuals obtained
  after analyzing the simulated data in GR (which means after the fit
  of the different initial conditions using a GR model) and represents the expected physical signal.}
\label{figCass_chi1}
\end{figure}

On \fref{figCass_gamma}, we can see that the signal due to the
conjunction is not absorbed at all by the fit of the initial
conditions which nevertheless absorbs modulations between the
conjunctions. Another example is given in \fref{figCass_chi1}
where the influence of a linear term in the spatial part of the
metric is shown. As can be seen, in this case, the fit of the
initial conditions absorbs a big part of the signal (the signal is
absorbed by a factor 100).

\begin{figure}[bt]
\begin{center}
\includegraphics[width=0.7\textwidth]{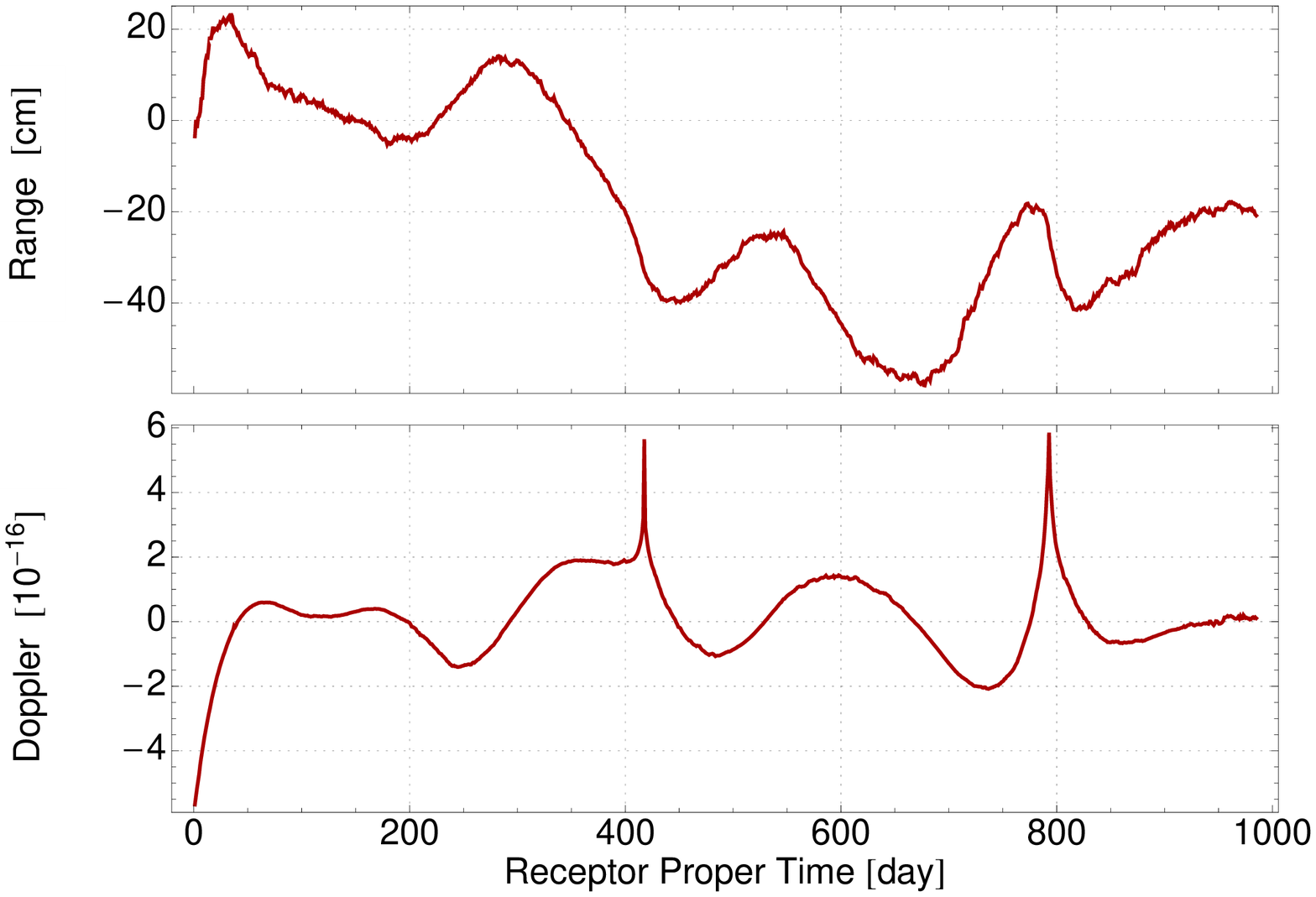}
\end{center}
\caption{Representation of the difference of the Range (top) and Doppler (bottom) residuals between residuals computed with and without the presence of Jupiter for an alternative theory with $\chi_1= 10^{-21}\ m^{-1}$ (and all other PEG parameters vanishing). The order of magnitude has to be compared with the order of magnitude of the residuals (green continuous lines) presented in \fref{figCass_chi1}: the presence of Jupiter changes the residuals obtained by only few~\%.}
\label{figCass_Jup}
\end{figure}

In order to illustrate that the simplified situation considered here is sufficient to obtain correct residual templates, we performed simulations by adding Jupiter. \fref{figCass_Jup} represents the difference between the residuals obtained by taking into account Jupiter and without the giant planet in the PEG simulation and in the GR analysis. Comparing the order of magnitude of \fref{figCass_chi1} and \fref{figCass_Jup}, we can see that the addition of Jupiter does not change the residuals by more than roughly 5~\%.

\begin{figure*}[htb]
\begin{center}
\includegraphics[width=0.49\textwidth]{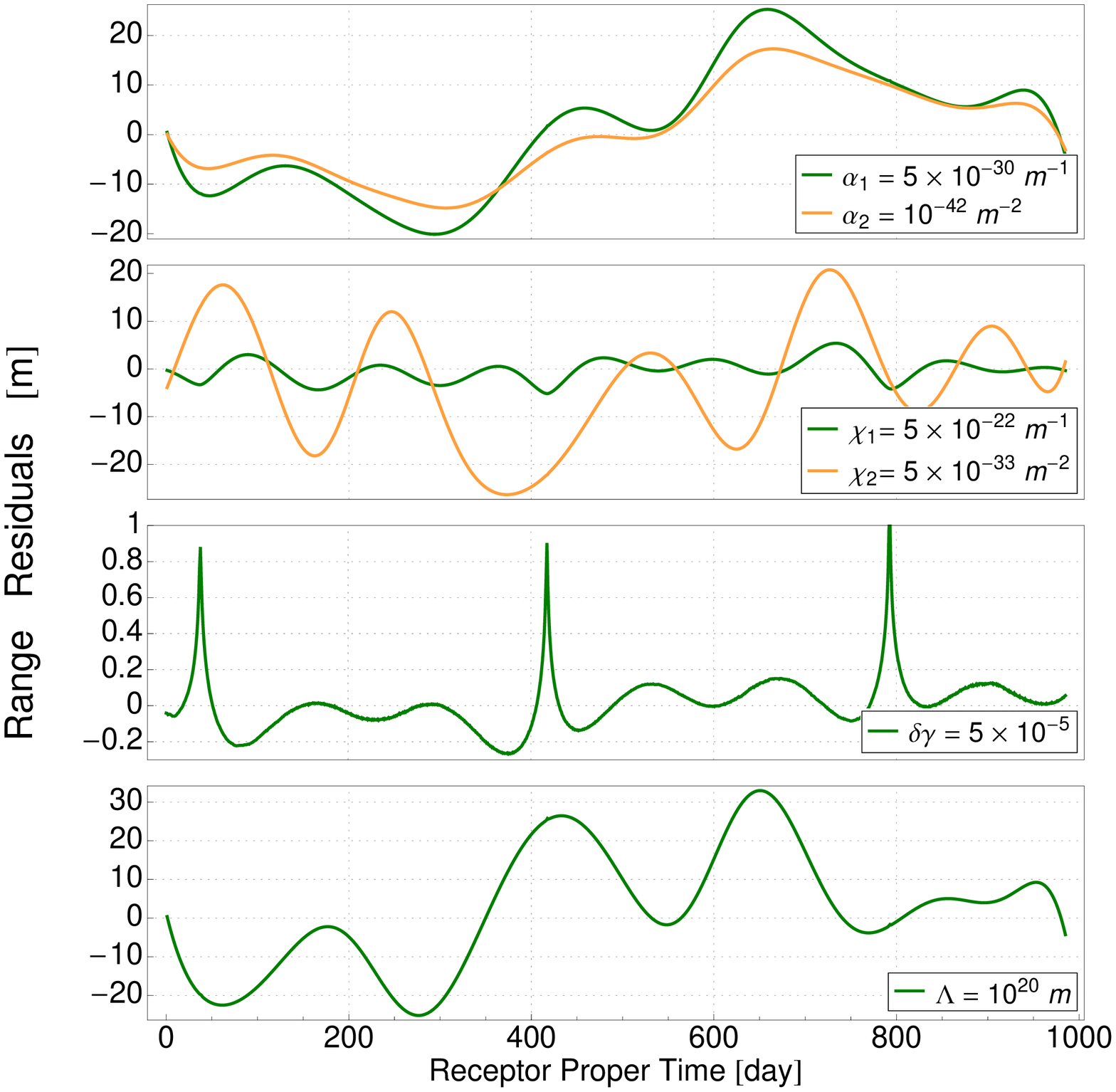}\hfill
\includegraphics[width=0.49\textwidth]{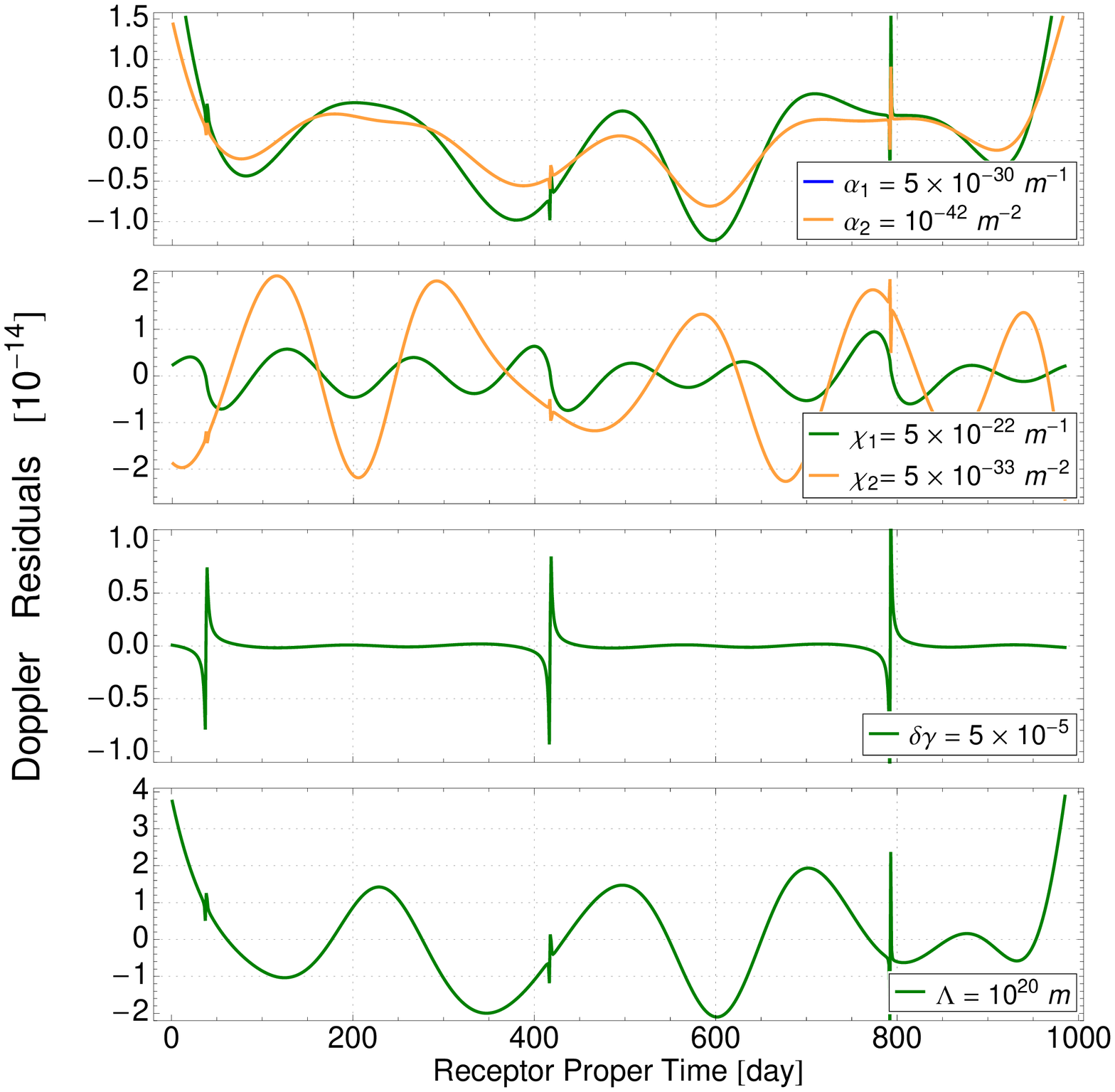}
\end{center}
\caption{Representation of the Range (left) and Doppler (right) residuals for all the PEG parameters considered. These residuals are anomalous signals produced if the gravitation theory is PEG but if the data were analyzed in GR. These templates can be searched in residuals of an analysis of real data.}
\label{figRes}                                                                        
\end{figure*}

\begin{figure}[htb]
\begin{center}
 \includegraphics[width=0.32\textwidth]{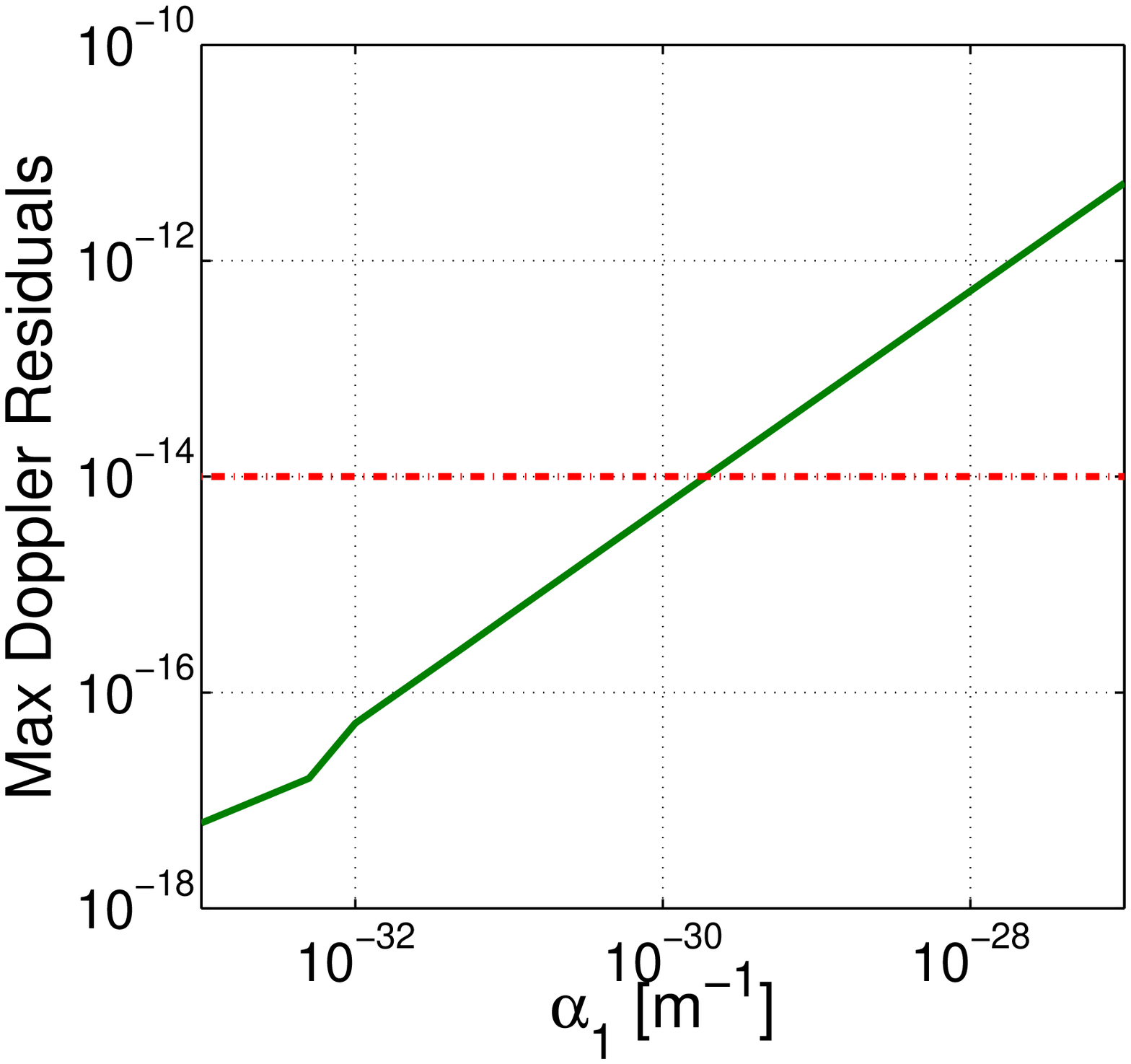} \hfill
 \includegraphics[width=0.32\textwidth]{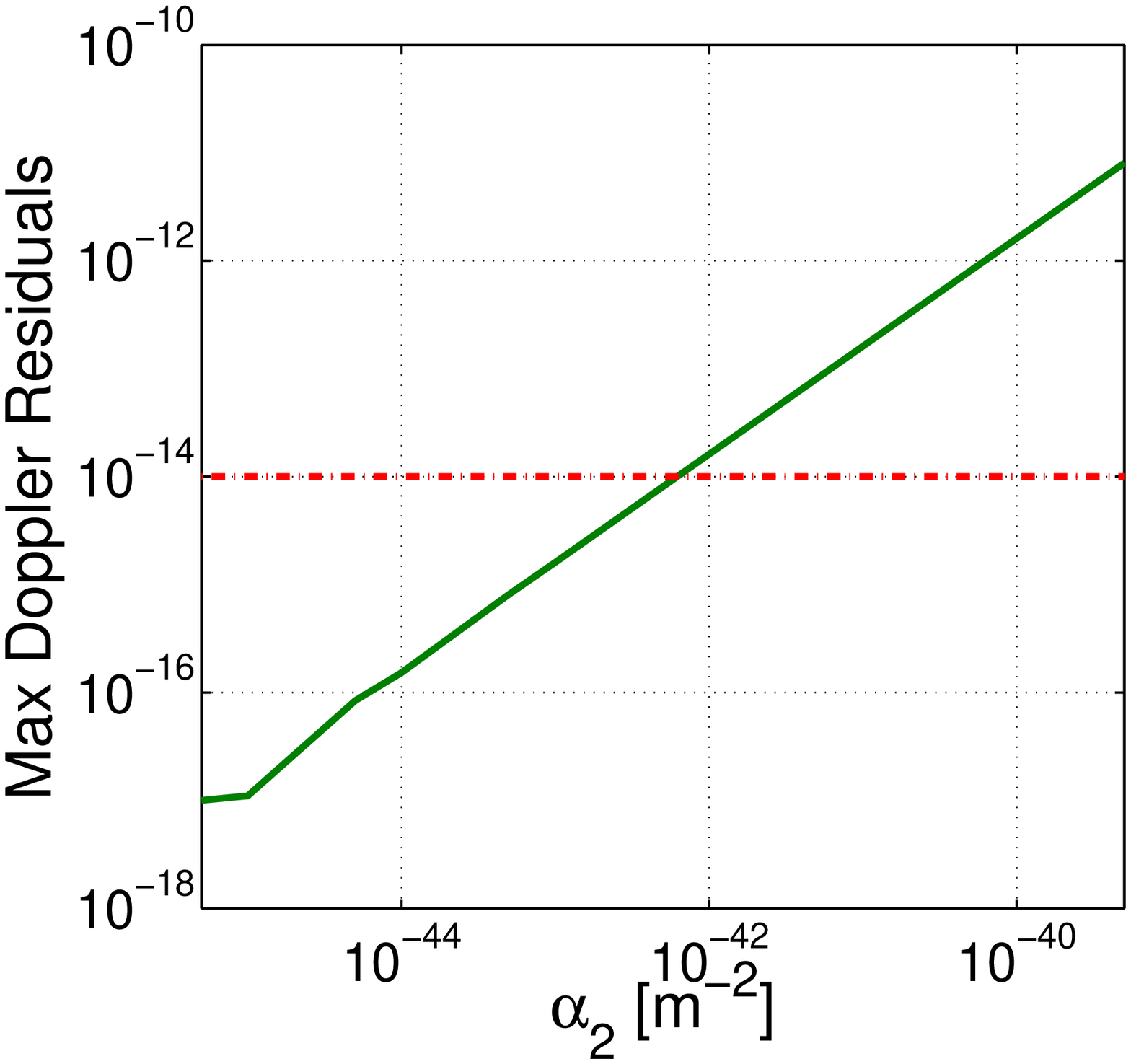}\hfill
 \includegraphics[width=0.32\textwidth]{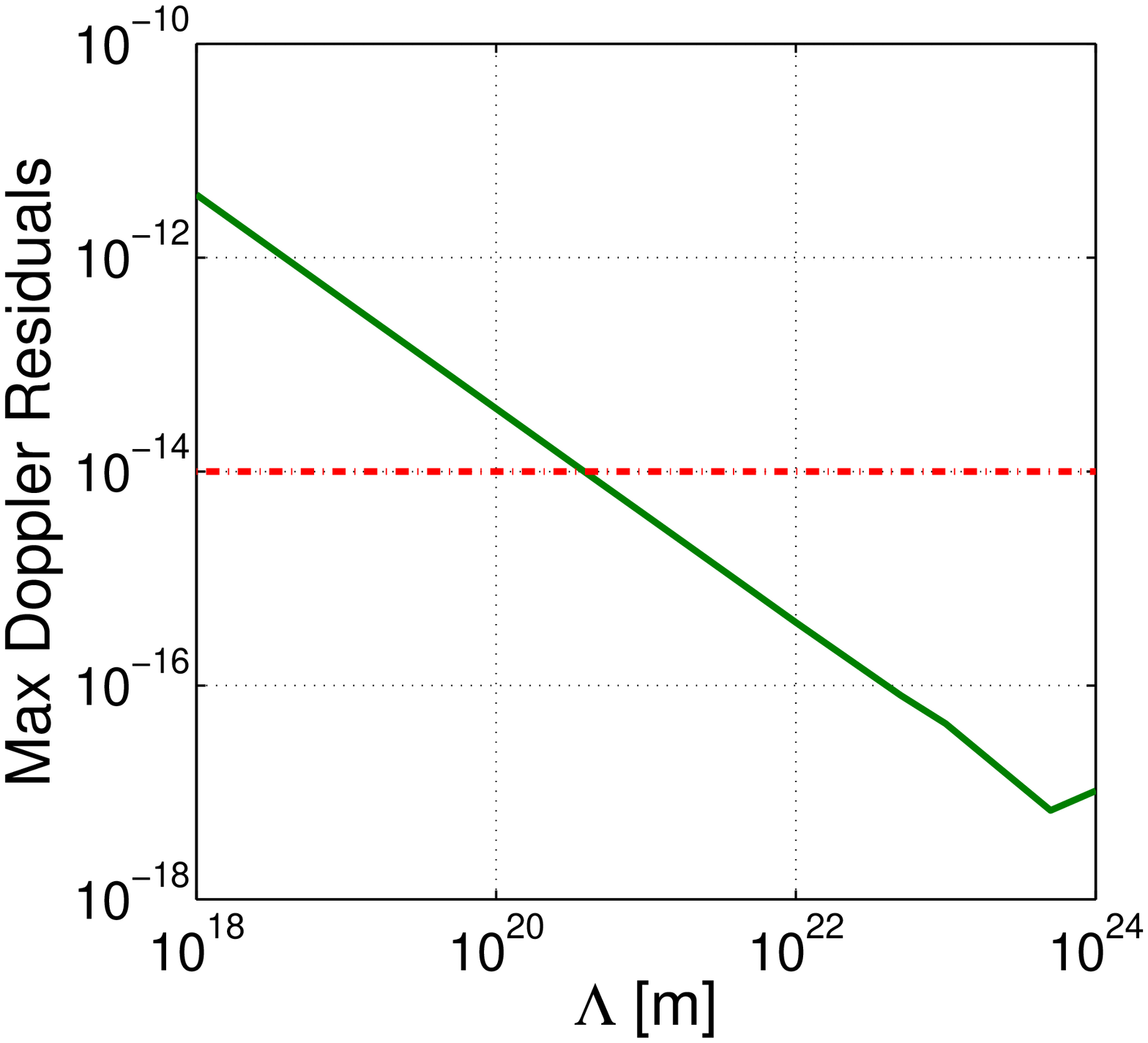}
 \includegraphics[width=0.32\textwidth]{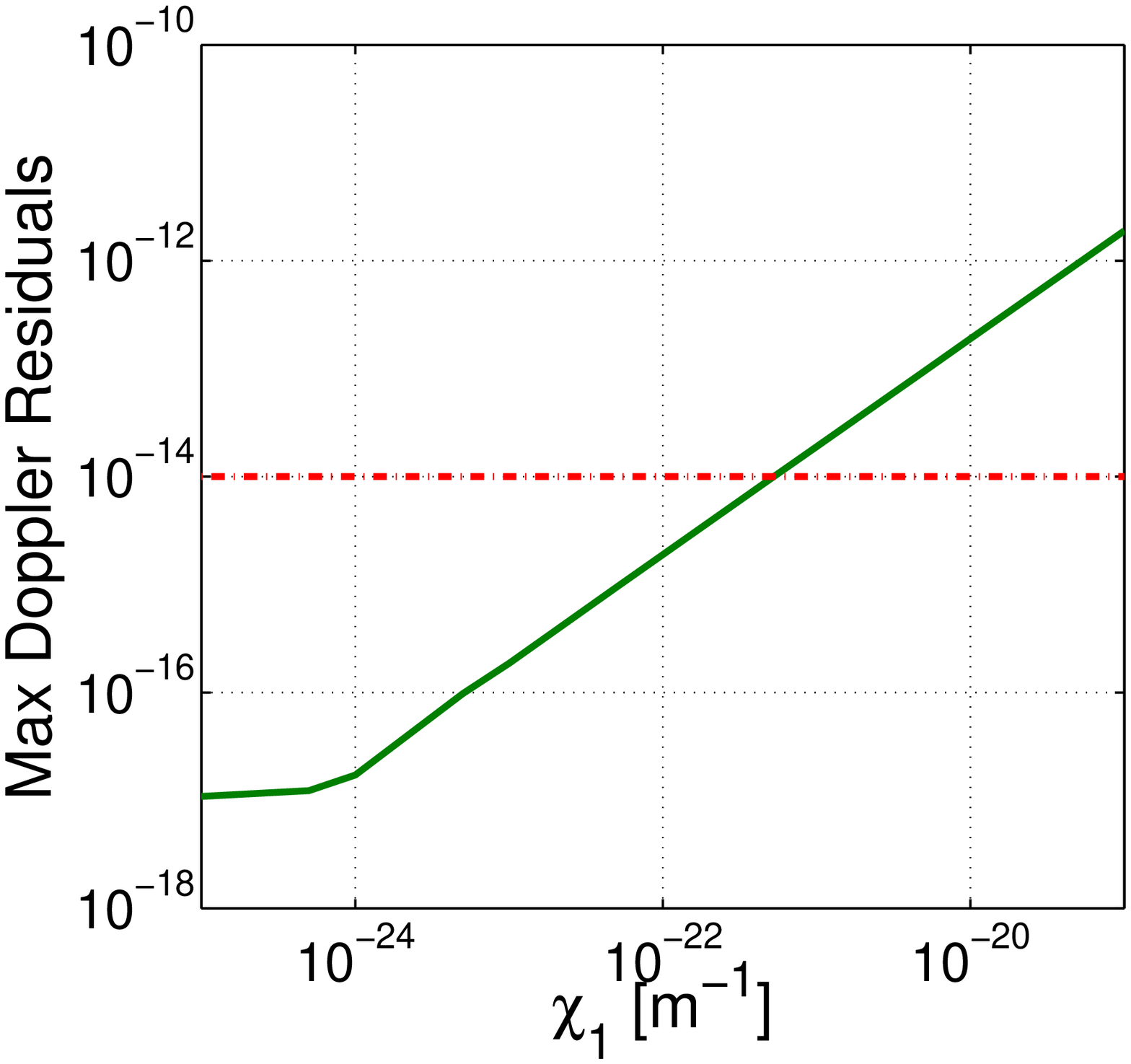}\hfill
 \includegraphics[width=0.32\textwidth]{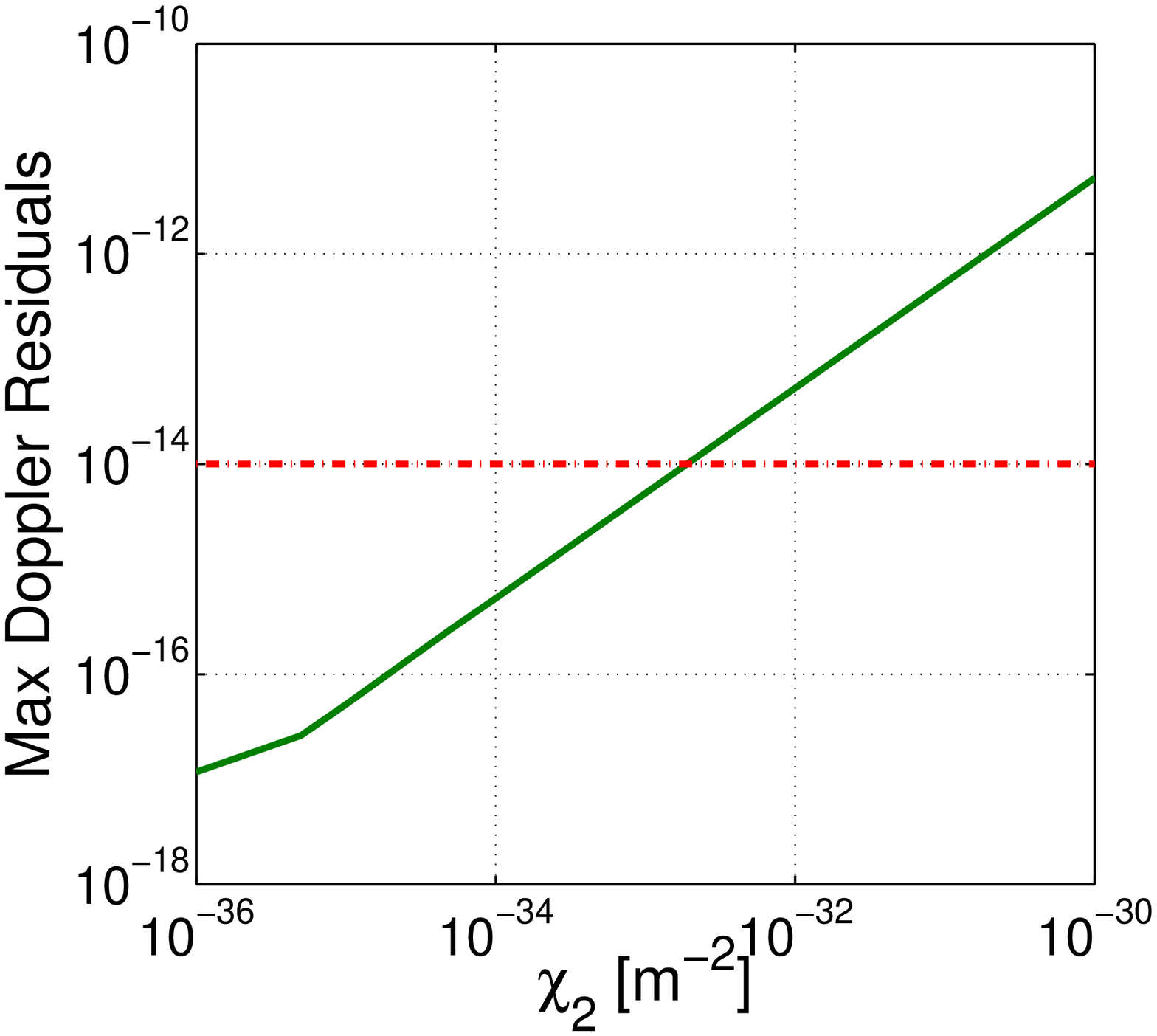}     \hfill
 \includegraphics[width=0.32\textwidth]{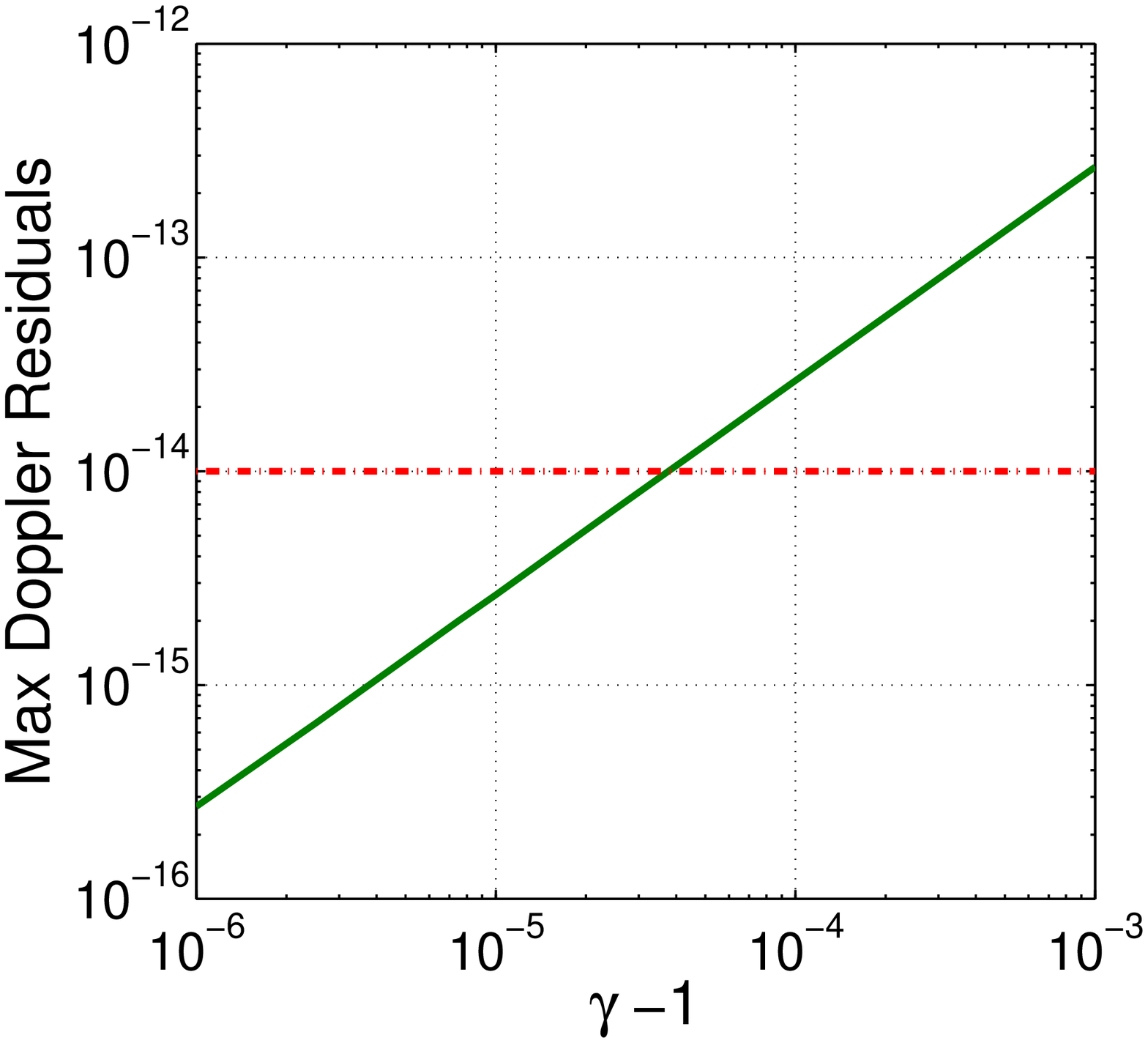}
\end{center}
\caption{Representation of the maximal Doppler signal due to PEG
  theory (parameterized by the 6 parameters as indicated by the expansion
  (\ref{dphin}-\ref{dphip}) for the Cassini mission between Jupiter
  and Saturn. The green (continuous) lines represent the
  maximum residuals obtained after analyzing the simulated data in GR
  (i.e. after the fit of the initial conditions). The red (dashed) lines represent the assumed Cassini precision. Each subfigure represents the maximum of the Doppler residuals with respect to one PEG parameter (with all other PEG parameters vanishing).}
\label{figCass_PEG}
\end{figure}

\Fref{figRes} represents the residuals obtained for each PEG parameter considered. These are the signatures that need to be searched for systematically in the residuals of the GR analysis of real satellite data. 
\Fref{figCass_PEG} summarizes all the simulations done. These
figures represent the maximal difference between the Doppler generated
in different PEG theories and the Doppler generated in GR. The
different PEG theories are characterized by the values of their six
parameters $\alpha_1$, $\alpha_2$, $\Lambda$, $\chi_1$, $\chi_2$,
$\delta\gamma$. The blue (dash-dot) lines represent the maximal
differences between simulations in PEG theory and simulations in GR
with the same initial conditions. The green (continuous) lines
represent the maximal residuals obtained after analyzing the signal
generated in PEG theory in GR (i.e. after the fit of the initial
conditions). More precisely, the green lines represent the maximal
Doppler signal that we expect to see in the Cassini residuals if the
theory of gravity is PEG theory with the considered
parameters. Assuming a Cassini Doppler precision of roughly $10^{-14}$ (represented by the red (dashed)
curves on \fref{figCass_PEG}), we derive the order of magnitude of the uncertainties one would obtain on the parameters of the theory when carrying out a search on the residuals from a complete GR analysis of real data. These uncertainties  are given in \tref{tab:pegconst}. 
Were one of these six PEG parameters larger than the value indicated in the
table, a signal larger than Cassini precision would appear in the Doppler residuals (under the assumption that the signal is not completely absorbed by a fit of additional parameters of effects that are not considered here (eg. thermal radiation, solar pressure,\dots). . 
 Clearly, a complete realistic data analysis would be necessary if an anomalous signal showed up with the right signature in the data. Then, a refined treatment taking into
account the temporal signature of the signals and the spectral
signature would be necessary. The boundary obtained on $\gamma$ is of
the same order of magnitude as the one obtained by Bertotti et
al~\cite{bertotti:2003uq} with the analysis of the real data ($
\gamma-1 =(2.1\pm2.3)\times 10^{-5}$). The other values are completely
new. It is also interesting to note that the boundary value on
$\Lambda$ is of the order of $10\ kpc$ which corresponds to the
galactic distance.
\begin{table}[ht]\centering
\begin{tabular}{|c|c|}
\hline
 &  smaller than \\
\hline
$\alpha_1$ & $1.9 \times 10^{-30} \ m^{-1} =5.9 \times 10^{-11} \ kpc^{-1}$ \\
$\alpha_2$ & $6.2 \times 10^{-43} \ m^{-2}=5.9\times 10^{-4}\ \ kpc^{-2}$\\
$\Lambda^{-1}$ & $2.6 \times 10^{-21} \ m^{-1}=0.08 \ kpc^{-1}$\\
$\chi_1$ & $5.3 \times 10^{-22} \ m^{-1}=0.02 \ kpc^{-1}$\\
$\chi_2$ & $1.9 \times 10^{-33} \ m^{-2}=1.8 \times 10^6 \ kpc^{-2}$ \\
$\delta\gamma$ &  $3.7 \times 10^{-5}$ \\
\hline
\end{tabular}
\caption{Estimate of the uncertainties on the six PEG parameters considered in this
paper obtainable in a complete analysis with real data. These values are obtained by requesting the maximal
residuals generated by the alternative theory to be smaller than
the assumed Cassini Doppler precision. In the computation of these values, we vary only one parameter at a time
independently of the others (set to zero). } \label{tab:pegconst}
\end{table}

\subsection{Simulations with MOND external field effect}
\begin{figure}[htb]
\begin{center}
\includegraphics[width=0.7\textwidth]{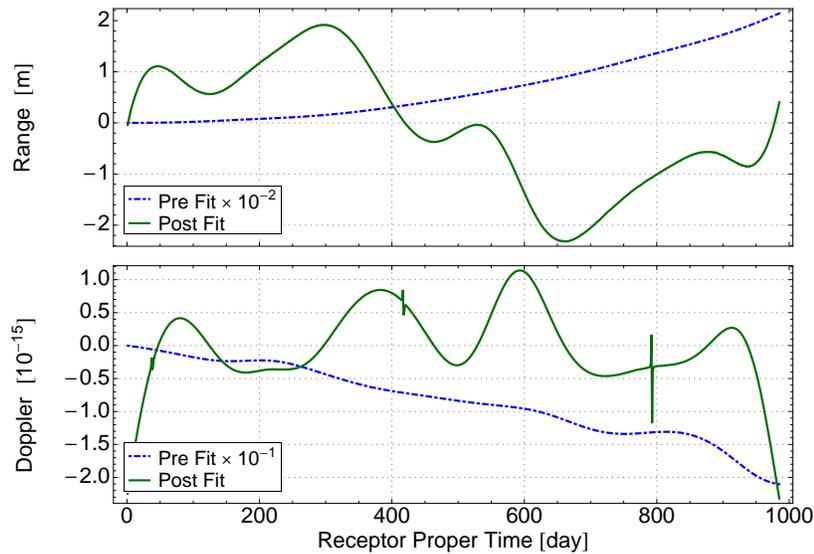}
\end{center}
\caption{Representation of the Range (top) and Doppler (bottom) signals
  due to the MOND external field effect with $Q_2=4.1 \times 10^{-26} \
  s^{-2}$. The blue (dash-dot) line is the difference between a
  simulation in the alternative theory and a simulation in GR (with the
  same initial conditions) and is for illustration purposes only (see text at the beginning of \sref{sec:results}). The green (continuous) line is the residuals
  obtained after analyzing the simulated data in GR (which means after
  the fit of the different initial conditions) and represents the expected physical signal.}
\label{figCass_mond}
\end{figure}

To analyze the influence of the MOND EFE, we use the metric
(\ref{pegmetric1}-\ref{pegmetric2})  with the modifications of the potentials given
by (\ref{dphimond}). \Fref{figCass_mond} represents the Range
and Doppler difference between a simulation including MOND EFE
(with the maximum value of the quadrupole allowed $Q_2=4.1 \times
10^{-26} \ s^{-2}$) and a pure GR simulation. The residuals are too small to be
detected with the considered arc of the Cassini mission. For this
reason, the Cassini radioscience experiment (when the cruise was
between Jupiter and Saturn) is not sensitive enough to the MOND
EFE. However, it may well be visible with a longer arc of data, or
when combining arcs from several spacecraft and possibly planetary
observations. This will be the subject of future work. Finally, let us recall that the MOND EFE is very well
constrained by the planetary ephemerides as indicated in
\sref{sec:outline}.

\section{Conclusion}
It is still an important challenge to test GR in the Solar System. Here, we focussed on the possibility to test GR with
radioscience measurements. As emphasized in the introduction, it
is essential to test GR in regimes not yet explored. This means
looking either for deviations smaller than the current constraints
(for example on the Post Newtonian Parameters) or for deviations
in a more general framework than the ones used until today (mainly
the PPN and the fifth force frameworks).

The work presented in this paper part of a project whose goal is to
scan data from solar system observations for eventual violations of
GR. Once completed, the full procedure will consist of four steps:
simulations of observables in an alternative theory of gravity
considering a simplified situation where only elements producing
significant deviations from GR are simulated; analysis of these
simulated observables using the usual procedure in GR to obtain the
incompressible residuals due to the theory considered~\cite{zarrouati:1987fk}; analysis of the
real data using standard procedure in GR (including all known
systematic effects); systematical search of the residuals of step (iii)
using the template obtained in step (ii). In this paper, we have focussed
on the two first steps. In particular, concerning the first step, we
have presented a software aiming at simulating Range/Doppler
observables directly from the space-time metric. This tool makes it
easy to change the theory of gravity (the only thing to change is the
metric). The method used to simulate Range and Doppler from the metric
has been presented into detail.  Moreover, concerning the second step
of the procedure, we have used a software doing a simplified version
of the traditional analysis in GR by means of a least-square fit of
the different initial conditions involved in the problem.

While being very general, this approach has some limitations which are nevertheless justified since we are considering only the leading terms in the deviation from GR. Therefore, the gravitating bodies (Sun and Earth) are approximated as point masses. Note that if necessary, multipolar expansions could be taken into account. Nevertheless, the impact due to a modification of the gravitation theory on the multipolar expansion should be negligible (the effect of an alternative theory on the monopole term is already expected to be small). Considering point masses also means we suppose the observer to be located at the center of the Earth. The second simplification done is to neglect effects coming from a hypothetical violation of the Strong Equivalence Principle (SEP). A violation of the SEP implies a Nordtvedt term (parameterized by the Nordtvedt parameter $\eta$) in the equations of motion~\cite{nordtvedt:1968uq,nordtvedt:1968ys} which can be added in the software if necessary. Finally, an implicit assumption done using the least-squares fit is that the error distribution of the measurements to be Gaussian.
 
Our results correspond to simulations in two alternative theories of gravity: Post-Einsteinian Gravity and the External Field Effect due to a MOND theory of gravity. The simulations have been performed for the Cassini spacecraft during its cruise between Jupiter and Saturn. The Range and Doppler residuals due to these theories have been presented in \fref{figRes}. These residuals furnished templates for signatures that can be searched for in real data analysis. The parameters uncertainties reachable in a complete analysis with real data have been estimated and are given in \tref{tab:pegconst}. 

The External Field Effect due to a MOND theory of gravity on the considered arc of the Cassini mission is just too small to be observed. This arc can
not give a significant constraint on the MOND theory, astrometric data based on perihelia precessions giving better constraints~\cite{fienga:2011qf}.

Let us summarize the innovative points of this work. The approach followed by deriving radioscience signals from the space-time metric is very general and makes it possible to obtain observables in alternative theories of gravity independently from the coordinate systems used and independently from any exterior data treated in GR (for example without any reference to ephemerides computed in GR). The fit of the initial conditions which is quite often forgotten in the analysis of anomalies can reduce the deviations produced in the observables quite significantly depending on the theory considered. A first crude limit on PEG parameters can be derived by using the Cassini spacecraft. Finally, this software can be used to test other alternative theories of gravity.

The method presented here can be generalized to non metric theories provided the equations of motion of massive bodies and the equations of light propagation are known. In general, these equations can be derived from the field equations. Simulations in nearly every alternative theory of gravity (for example Standard Model Extension, TeVeS, \dots) can thus be made.  Another perspective is to extend the software to simulate observations related to angular measurements (VLBI, position of star in the sky, position of planets) which constitutes the other type of measurement done in the Solar System. Such an extension will allow one to simulate all the observations done in the Solar System in any metric theory of gravity and to derive the expected signals in the residuals when analyzing those observables in GR.

Finally, radioscience data from existing and future space missions could be analyzed to derive more precise constraints on alternative theories of gravity.

\ack
A. Hees is supported by an FRS-FNRS (Belgian Fund for Scientific Research) Research Fellowship. Numerical simulations were made on the local computing resources (cluster URBM-SysDyn) at the University of Namur (FUNDP).

\section*{References}    

\bibliographystyle{unsrt}
\bibliography{../../../Documents/byMe/biblio}

\end{document}